\documentclass[review]{elsarticle}

\usepackage{hyperref}
\usepackage{xcolor}
\usepackage{graphicx}
\usepackage{epstopdf, epsfig}
\usepackage[export]{adjustbox}
\usepackage{caption}
\usepackage{subcaption}
\usepackage[english]{babel}
\usepackage[letterpaper,top=2cm,bottom=2cm,left=3cm,right=3cm,marginparwidth=1.75cm]{geometry}
\usepackage{amsmath}
\usepackage{MnSymbol}%
\usepackage{wasysym}%
\usepackage{graphicx}
\usepackage{arydshln}

\usepackage{color}
\usepackage{ulem}

\usepackage{mathMacros}

\journal{Journal of Non-Newtonian fluid Mechanics}

\newcommand{\Pab}[1]{{\color{blue}#1}}

\newcommand{\vonK}{von K\'{a}rm\'{a}n }

\begin{document}

\begin{frontmatter}

\title{Elastoviscoplastic fluid flow past a circular cylinder}

\author[label1]{S. Parvar\corref{mycorrespondingauthor}}
\author[label1]{K. T. Iqbal}
\author[label1]{M. N. Ardekani}
\author[label1,label2]{L. Brandt}
\author[label1]{O. Tammisola}

\address[label1]{SeRC and FLOW, Engineering mechanics, KTH Royal Institute of Technology, SE-10044 Stockholm, Sweden.}
\address[label2]{Department of Energy and Process Engineering, Norwegian University of Science and Technology (NTNU), Trondheim, Norway}

\cortext[mycorrespondingauthor]{Corresponding author \\ Email address: parvar@kth,se, \href{https://www.researchgate.net/profile/Saeed-Parvar}{(S. Parvar)}.}

\begin{abstract}
{The combined effect of fluid elasticity and yield-stress on the flow past a circular cylinder is studied by two-dimensional direct numerical simulation. We analyze the effects of yield-stress, elasticity, shear-thinning, and shear-thickening on the wake characteristics using the Saramito constitutive model. 
The elastoviscoplastic (EVP) wake flow is studied at a moderate Reynolds number ($Re= 100$) where two-dimensional vortex shedding occurs in the Newtonian case. We find that in the shear-thinning elastoviscoplastic flow, when yield stress increases, the drag coefficient and root mean square of the lift coefficient both decrease, while the length of the recirculation bubble $L_{RB}$ increases. These changes indicate that the wake oscillation amplitude decreases with an increasing yield stress. For shear-thickening however, the drag coefficient $C_{D}$ increases at a large Bingham number, and the wake becomes chaotic. The comparison of viscoelastic fluid and EVP fluid reveals that the polymer stresses, $\trace{\btau^p}$, decay considerably less downstream of the cylinder in the EVP case, indicating that significant stresses persist at large distances. We observe that shear-thinning competes with elastic and yield stresses and counteracts their effect, while shear-thickening enhances elastic and yield stress effects, so that the flow pattern can change from periodic to a chaotic flow.}
\end{abstract}

\begin{keyword}
{Cylinder, Vortex Shedding, Saramito Herschel-Bulkley constitutive equation, Elastoviscoplastic fluids.}
\end{keyword}
\end{frontmatter}


\section{Introduction}\label{section:introduction}

{The flow around a circular cylinder is one of the canonical flow cases in fluid dynamics. Despite its simple geometry, it displays complex flow patterns due to the interaction of a shear layer, a separation region, and a wake. The onset and character of vortex shedding around the cylinder in a Newtonian fluid has been studied by numerous authors, including also confinement effects \cite{Williamson1988,Williamson1996,Posdziech2007,Rajani2009,Constant2017,Qu2013,Juniper2011,Tammisola2011}. Even Non-Newtonian cylinder flows have attracted plenty of attention due to their relevance for natural and industrial flows through porous materials, in geological applications, lab-on-the-chip devices, and human biology, to mention few examples \cite{Balmforth2014,Mossaz2010,Saramito2016}. It can be mentioned that \cite{Iman2012} studied the linear global stability of inelastic Carreau fluids, and found the critical Reynolds number for onset of wake instability decreased all the way down to $Re_{cr}=3$ for shear-thinning fluids, while shear-thickening had a stabilizing effect. Both effects were attributed to changes in the mean flow wake structure. 

Next, let us now consider the role that elasticity plays in unsteady shear flows. The addition of a few parts per million of polymers makes the flow viscoelastic, which can result in significant changes in the flow features compared to Newtonian flows. Turbulent drag reduction in viscoelastic fluids was reported by Toms \cite{Toms1948} for the first time, and this effect has been studied by numerous authors in wall-bounded turbulent flows, such as pipe, channel and boundary layer flow \cite{Virk1967,PARVAR20211}, and free shear flows, such as jets \cite{PARVAR20201,PARVAR20202,PARVAR20212} and mixing layers \cite{PARVAR20222}. Viscoelastic drag reduction in turbulent flows  has been the subject of many reviews, see Lumley \cite{Lumley1969}, Virk \cite{Virk1971}, and White \cite{White2003}. It was more recently discovered that the presence of polymers brings a creeping flow to a chaotic regime, the so-called elastic turbulence (ET) \cite{Steinberg2021}. This elastic turbulence is driven by the nonlinear elastic stresses that arise from stretched polymer molecules in a highly elastic solution, i.e.\ is characterised by a high Weissenberg number ($Wi>>1$) and a high Elasticity number $EI=(Wi/Re)>>1$. In the present study, with $Re=100$, inertia is significant, resulting in a low elasticity number ($0\leq EI \leq 0.04$), and hence elastic turbulence is not expected a priori. In the following, the studies of the viscoelastic flow past a circular cylinder are reviewed.

The first studies were the experiments by Smith \etal \cite{Smith1967} and 
James and Acosta \cite{James1975} and the numerical simulations based on the generalized Maxwell model by Mizushina and Usui\cite{Mizushina1975}. Later, Chilcott and Rallison \cite{Chilcott1988} proposed the FENE-CR constitutive equations to study the flow of a  dilute polymer solution past a cylinder at a low Reynolds number. These authors observed that the polymer molecules are highly extended at a large $Wi$ number in the area close to the rear stagnation points of the cylinder. McKinley et al. \cite{McKinley1993} investigated experimentally the creeping flow of a highly elastic fluid past an object placed inside a channel, and how $Wi$ affects the flow features and the formation of a vortex street. These authors reported that increasing $Wi$ up to moderate values ($Wi=1.3$) led to a change from a steady two-dimensional flow to a steady three-dimensional flow with a spanwise periodic cellular structure. Further increasing elasticity to $Wi \ge 1.85$ caused the flow to become unsteady and chaotic (identified by non-linear fluctuation and complex spatial structure). Oliveira \cite{Oliveira2001} used the FENE-CR model to study the effect of shear-thinning and elasticity on the flow past a cylinder. This author noticed that the vortex shedding frequency and mean drag coefficient decreased with increasing $Wi$ , while the length of the recirculation bubble increased. Coelho and Pinho \cite{Coelho2003I,Coelho2003II} also conducted experimental studies to investigate the effect of elasticity and shear-thinning on vortex shedding. They reported that shear-thinning decreases the boundary-layer thickness and increases the shedding frequency, i.e.\ the Strouhal number ($St$), while increasing $Wi$ decreases the $St$ number. They also stated that if shear-thinning dominates over elasticity, the $St$ number increases. Oliveira and Miranda \cite{Oliveira2005} performed several numerical simulations of a FENE-CR fluid past a confined cylinder and reported that the creeping flow becomes unsteady and periodic at a large $Wi$. They observed a small recirculation bubble attached to the cylinder downstream of the stagnation point. The presence of a recirculation bubble caused the drag coefficient to increase and to vary sinusoidally, unlike in the steady-state situation without the bubble. They also reported that the bubble length oscillates in time. 

Xiong et al. \cite{Xiong2010} conducted direct numerical simulations of an Oldroyd-B fluid flowing past a cylinder. Their findings indicate that for Reynolds numbers less than 40, the drag coefficient increases with the Weissenberg number. However, for Reynolds numbers greater than 40, the effect of the Weissenberg number on the drag coefficient is non-monotonic, meaning that it can either increase or decrease. This indicates that the changes in the drag at $Re>40$ are associated with changes in the vortex shedding. Richter et al. \cite{Richter2010} studied the viscoelastic flow past a cylinder for two different Reynolds number $Re=100, 300$, using the FENE-P closure for the polymer stresses, as generally used for the numerical simulation of a dilute polymer solution. These authors studied the effect of viscoelasticity on the transition to three-dimensionality in the cylinder wake at varying Reynolds numbers and maximum extensibility of the polymers. 
They explained that at $Re=100~$ the polymeric stresses lengthen the recirculation region, increase the average drag force, and weaken the von Kármán instability compared with a Newtonian fluid, or compared with a polymer solution with a lower maximum extensibility. At a larger Reynolds number $Re=300$ and when $Wi$ is large enough, the three-dimensional Newtonian mode B instability is suppressed, and all three-dimensional structures disappeared due to higher polymer extensibility.
Sahin and Atalik \cite{Sahin2019} carried out two-dimensional finite element simulations to study the flow of an inelastic power-law fluid, viscoelastic Oldroyd-B fluid, Giesekus fluid, and FENE-P fluid past a circular cylinder in the periodic vortex shedding regime. They observed that shear-thinning reduces the drag coefficient, the recirculation bubble length, and the separation angle, while it increases the frequency of vortex shedding in an inelastic fluid. However, elasticity plays the opposite role; it increases the drag coefficient and the length of the recirculation bubble while it decreases $St$. In addition, these authors show that increasing elasticity decreases the separation angle, similarly to the shear-thinning effect.

Recently, Peng et al. \cite{Peng2021} studied the combined effects of the shear-thinning and Weissenberg number on the two-dimensional vortex shedding of a viscoelastic fluid flow past a circular cylinder at $Re=100$. The Giesekus model with the log-conformation formulation was used to perform the simulations. These authors report that pure shear-thinning can reduce the apparent viscosity and induce an inertial instability near the cylinder surface, while the extensional viscosity due to the presence of the polymer molecules suppresses this instability. However, the combination of shear-thinning and elasticity lengthens the recirculation bubble and reduces the lift coefficient fluctuations, $C_{L_{rms}}$, and the Strouhal number, $St$. They also observed an elastic instability at very large $Wi$ which caused a large flow oscillation in the vicinity of the upstream stagnation point. Furthermore, both strong elasticity and strong shear-thinning increase the drag coefficient; nevertheless at a low $Wi$ number of a shear-thinning solution, drag may reduce.

Minaeian et al. \cite{Minaeian2022} used the Rheofoam toolbox of OpenFOAM \cite{RheoTool} to study the effect of viscoelasticity on the onset of vortex shedding in a high concentration polymer solution (with a high polymer/solvent viscosity ratio). The Phan–Thien Tanner (PTT) model was utilized with a finite volume and log-conformation method to solve the governing equations. They observed that at a high concentration solution of PTT fluids, the elasticity effect varies depending on the value of the Elasticity number. At low $EI$, the flow becomes unstable at larger $Re$, while at large $EI$, the flow became unstable at lower $Re$ in comparison with a Newtonian flow. The same authors showed that $Re_{cr}$ decreases linearly with ($1-\beta_{s}$) (where $\beta_{s}$ denotes the ratio of solvent viscosity to total viscosity), and increases with the extensional viscosity. Viscoelasticity extends the vortices and may reduce their sizes, and the vortex shedding frequency reduces at a larger extensional viscosity.

Studies of viscoplastic effects on the vortex shedding around a cylinder are very scarce. Mossaz et al. \cite{Mossaz2010} studied the viscoplastic flow past a cylinder in an unconfined domain, using a regularized version of the Herschel-Bulkley model. They considered the effects of the power-law index $n$ in the range of shear-thinning and shear-thickening ($0.3 \leq n \leq 1.8$), and the yield stress, characterized by the Bingham number ($0 \leq Bn \leq 10$), on vortex shedding behind a circular cylinder and reported that increasing $Bn$ increases the critical $Re_{cr}$ and $St$ numbers.} At constant $Re$, yield stress decreases the vortex shedding frequency. Kanaris et al. \cite{Kanaris2015} carried out three-dimensional direct numerical simulations of a viscoplastic flow around a confined circular cylinder using the regularized Bingham model in the regime where the wake transitions to three-dimensionality ($0 \leq Bn \leq 5$ and ($150 \leq Re \leq 600$)). They observed that the critical Reynolds number for the onset of the three-dimensional ﬂow regime, $Re_{cr}$, increases linearly with the Bingham number. The same authors also reported substantial changes in the flow structure downstream of the cylinder.

In the present work, the elastoviscoplastic (EVP) flow past a circular cylinder is studied. The Saramito elastoviscoplastic constitutive equation \cite{Saramito2009} is solved in a fully coupled manner with the incompressible Navier-Stokes equations. The log-conformation approach is utilized to avoid the well-known high Weissenberg number problem (HWNP) \cite{Fattal2004,Fattal2005}. 
Extensive simulations are performed to systematically study the effect of the rheological characteristics on the vortex shedding and fluid structures at various $Wi$ and $Bn$ numbers for dilute and concentrate polymer solutions. To the best of our knowledge, this is the first study considering the combined effects of the yield stress and elasticity on a flow past a circular cylinder in the vortex shedding regime.

The paper is organized as follows: after introducing the flow of interest and the computational domain in \Secref{section:Flow_Setup}, the complete set of governing equations is explained in \Secref{section:Governing_Equation}. \Secref{section:Numerical_Algorithm} addresses the numerical algorithms utilized in the present work, i.e.\ the immersed boundary method (IBM) and the log-conformation tensor approach. \Secref{section:Results_Discussion} discusses the new finding of the present work, starting with the verification of the in-house code for standard and canonical Newtonian and non-Newtonian fluid problems. Finally, the conclusions of the present work are presented in \Secref{section:Conclusion}.

\section{Flow Setup}\label{section:Flow_Setup}
\begin{figure} 
\centering
\includegraphics[width=1.0\textwidth]{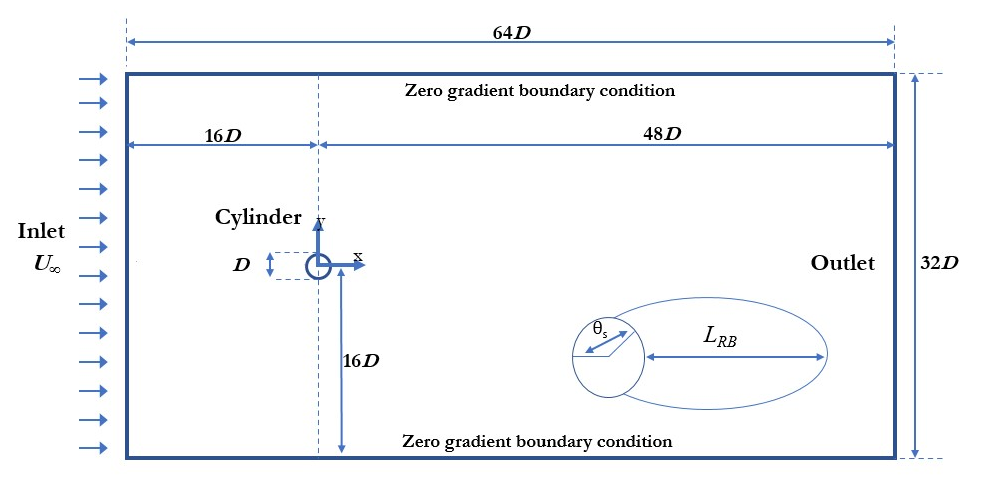}
\caption{\label{fig:image1}The schematic of flow past a cylinder, the recirculation bubble length $L_{RB}$ and the separation angle $\Theta_{s}$}
\end{figure}

\Figref{fig:image1} displays the schematics of the flow of interest in which a free stream velocity $U_{\infty}$ flows past an infinite stationary circular cylinder of diameter $D$, the flow direction, and the coordinate system. The origin of the domain is located at the center of the cylinder, as shown in \Figref{fig:image1}. The lengths are expressed in terms of the cylinder diameter $D$, the velocities with the free stream velocity $U_{\infty}$ and stresses are normalized using the inertial scaling \ie $\rho U_\infty^2$, where $\rho$ denotes the fluid density. Two attached symmetric closed eddies are formed just behind the cylinder and create a steady flow recirculating region, called the recirculation bubble. At around $Re=47-49$, the steady recirculation bubble behind the cylinder becomes unstable, and the well-known \vonK vortex street appears as two parallel rows of vortices \cite{Williamson1996,Giannetti2007}. In addition to vortex shedding, one of the main flow features is the recirculation bubble length, defined as the distance from the rear stagnation point to the end of the recirculation region, denoted by $L_{RB}$ in the present work. If the amplitude of the shed vortices increases, the Reynolds stresses increase and the length of the recirculation bubble decreases. The location of the separation point on the cylinder surface is another interesting feature, represented by the separation angle, $\Theta_{s}$, defined as the angle between the upstream stagnation point and the separation point. Both the definition of $L_{RB}$ and $\Theta_{s}$ are illustrated in \Figref{fig:image1}. We also note that the time-averaged $\Theta_{s}$ decreases by increasing $Re$ \cite{Qu2013,Jiang2020}.

\section{Governing Equation}\label{section:Governing_Equation}

We study the incompressible flow of a elastoviscoplastic liquid, modelled by the Saramito closure. To start with, we introduce
the relevant dimensionless numbers that govern the flow dynamics. Here, the reference viscosity is the total viscosity $\mu_{0} = \mu_s + \mu_p$, where $\mu_{s}$ and $\mu_{p}$ denote the solvent and polymer viscosities, respectively. 
The Reynolds number, $Re$, is defined as follows
\begin{equation}
    Re = \frac{\rho U_{\infty}D}{\mu_{0}},
    \label{eq:3.5}
\end{equation}

whereas the Weissenberg number, $Wi$, which is the ratio of elastic to viscous forces, is defined as
\begin{equation}
    Wi = \frac{\lambda U_{\infty}}{D},
    \label{eq:3.6}
\end{equation}
where $\lambda$ is the polymer relaxation time. 
The Bingham number, characterizing the ratio of the yield stress to viscous stresses,
\begin{equation}
    Bn = \frac{\tau_{y}}{\mu_{0}(U_{\infty}/D)},
    \label{eq:3.8}
\end{equation}

and, finally, the ratio of solvent viscosity, $\mu_{s}$, to total viscosity, $\mu_0$ (at the characteristic shear-rate of the flow), is denoted by 
\begin{equation}
    \beta_{s} = \frac{\mu_{s}}{\mu_{0}} = \frac{\mu_{s}}{\mu_{p}+\mu_{s}}.
    \label{eq:3.7}
\end{equation}
The governing equations consist of the continuity equation, momentum equation, and a constitutive equation for the EVP stress. The continuity and momentum equations read
\begin{align}
\label{eq:cont}
\divg{\tens{u}} = 0
\end{align}
\begin{align} 
\label{eq:mom}
\pddt{\tens{u}} + \advec{\tens{u}} = 
-\grad{p} + 
\frac{1}{Re}\divg{\btau^s} + 
\divg{\btau^p} + 
\tens{f}
\end{align}
where ${\tens{u}}={\tens{u}(\tens{x},t)}$, $p=p(\tens{x},t)$, and $\btau^p = \btau^p(\tens{x},t)$ denote the velocity vector, pressure, and EVP stress tensor, respectively. The solvent stress contribution is $\btau^s=2\beta_s \tens{S}(\tens{u})$, where
$\tens{S}(\tens{u})=(\grad{\tens{u}}+\trans{\grad{\tens{u}}})/2$ is the strain rate tensor. In \Eqref{eq:mom} the term $\tens{f}$ is a body force that is used to numerically impose the no-slip/no-penetration boundary conditions on the surface of the cylinder through the immersed boundary method, discussed in more detail in \Secref{section:IBM0}. As constitutive equation of the EVP fluid with shear rate-dependent viscosity, the Saramito constitutive equation \cite{Saramito2009} is used. It describes the evolution of the polymer stress, $\btau^p$, as follows:
\begin{equation}
Wi\, \ucd{\btau^p} +
F\btau^p -
\frac{2(1-\beta_s)}{Re}\tens{S}(\tens{u})=0,
\label{eq:SHB}
\end{equation}
where $F=\mathrm{max}\left(0,\dfrac{|\btau_d^p|-Bn/Re}{(2(1-\beta_s)/Re)^{1-n}|\btau_d^p|^n}\right)^{1/n}$. In \Eqref{eq:SHB}, the deviatoric stress tensor is $\btau_d^p = \btau^p - (\trace{\btau^p}/\trace{\tens{I}})\, \tens{I}$ (where $\tens{I}$ is the unit tensor), and the second invariant of $\btau_d^p$ is $|\btau_d^p| \equiv \sqrt{\btau_d^p : \btau_d^p /2}$. If $|\btau_d^p| \leq \tau_{y}$, the material deforms as a viscoelastic Kelvin-Voigt solid, whereas when $|\btau_d^p| > \tau_{y}$, the material flows as an elastoviscoplastic extension of the Herschel-Bulkley model. In \Eqref{eq:SHB}, $n$ is the power-law index quantifying the degree of shear-thickening $(n > 1)$ or shear-thinning $(n < 1)$; for $n=1$ the model reduces to the original Saramito model without shear-thinning \cite{Saramito2007}.
Finally, $\ucd{\btau^p}$ \footnote{It is noted that the upper-convected derivative in \Eqref{eq:ucd} is written by defining the velocity gradient such that $(\grad{\tens{u}})_{ij}=\partial u_i/\partial x_j$, \ie the same convention as in Gurtin \etal \cite{gurtin_fried_anand_2010} for instance.} is the upper-convected derivative 
of $\btau^p$, obtained as follows,
\begin{align}
\label{eq:ucd}
\ucd{\btau^p} \equiv \pddt{\btau^p} + \advec{\btau^p} - (\grad{\tens{u}})\, \btau^p - \btau^p\, \trans{\grad{\tens{u}}}.
\end{align}

The observable quantities used to interpret the results are the nondimensional shedding frequency, \textit{i.e.} the Strouhal number, $St=(D f_v)/u_{\infty}$ (where $f_{v}$ is the dimensional frequency), the drag coefficient $C_{D} = \frac{2 F_{D}}{\rho U_{\infty}^2 D}$ and the lift coefficient $C_{L} = \frac{2 F_{L}}{\rho U_{\infty}^2 D}$, where $ F_{D}$ and $ F_{L}$ are the drag and lift force per unit length. It is also to define the ratio between the root mean square (rms) of the drag to lift coefficients, $r=\frac{(C_{L_{rms}})}{(C_{D_{rms}})}$, which can be used as an indication of changes of the flow pattern.

\section{Numerical Algorithm}\label{section:Numerical_Algorithm}
\subsection{Immersed boundary method for representing the cylinder \label{section:IBM0}}
In the present study, the immersed boundary method (IBM) is used to represent the cylinder, still using a uniform computational grid, and hence exploit the possibility of utilizing efficient computational algorithms such as an efficient, highly-scalable, fast Fourier transform (FFT)-based pressure solver.

Here, the discrete forcing method is utilised to generate the cylinder surface, as in \cite{Izbassarov2018}. This method was proposed by Uhlmann \cite{Uhlmann2005} and then improved by Breugem et al. \cite{Breugem2012} to be second-order accurate in space by utilizing Luo et al.'s multi-direct forcing scheme \cite{Luo2007}. In this way, the simulation is performed on an uniform Cartesian and an Eulerian grid ($\Delta x = \Delta y$) with the solid cylinder surface reproduced by a uniformly distributed Lagrangian grid. 
Briefly, the algorithm is as follows. 
i) First velocity prediction $\tens{u}^{*}$ is calculated from Eq. \ref{eq:mom}; ii) the velocity is interpolated from Eulerian to Lagrangian grid on the cylinder surface in order to impose no-slip no-penetration at the cylinder surface (in this case assumed fixed). iii) Finally mass conservation is imposed by pressure projection. For more detail, we refer the reader to \cite{Breugem2012,Luo2007,Izbassarov2018}.

\subsection{The log-conformation tensor approach}\label{section:log0}
Many viscoelastic/elastoviscoplastic constitutive equations obey an evolution of the conformation tensor, $\tens{A}$, of the following form
\begin{align}
\label{eq:AeqnGeneral}
\pddt{\tens{A}} + \advec{\tens{A}} - \grad{\tens{u}}\,\tens{A} - \tens{A}\trans{\grad{\tens{u}}} = \frac{1}{\lambda} \tilde{g}(\tens{A})
\end{align}

where $\tilde{g}(\tens{A})$ is a model-specific function of $\tens{A}$. The log-conformation approach, proposed by Fattal and Kupferman \cite{Fattal2004,Fattal2005}, is used to handle the numerical instabilities appearing at high Weissenberg number, the well-known high Weissenberg number problem (HWNP). For the case of the Saramito model, $\btau^p$ and $\tens{A}$ are related as follows
\begin{equation}
\btau^p =\frac{1-\beta_s}{Wi}(\tens{A}-\tens{I}).
\label{eq:Kramers}
\end{equation}

The second-order central finite difference scheme on a uniform, staggered grid is used for Eqs. \eqref{eq:cont}, \eqref{eq:mom}, and \eqref{eq:SHB}, except for the advection term in \Eqref{eq:SHB} for which we use a fifth-order weighted essentially non-oscillatory (WENO) scheme \cite{Shu2009,Sugiyama2011}. The time integration is performed with a fractional-step, third-order explicit Runge–Kutta scheme \cite{Kim1985}, for all equations. For more detail, the reader is referred to \cite{Izbassarov2018,Izbassarov2021}.

\section{Code verification}
\label{section:verification}

\subsection{Numerical simulation details}\label{section:Computational_details}
The dimensions of the computational domain are: [-16 $D$, 48 $D$] × [-16 $D$, 16 $D$] in the streamwise ($x$) and cross-stream ($y$) and directions. To establish the inlet boundary condition, a uniform velocity profile is enforced, while a zero-gradient condition is set for the outlet. Confinement effects are considered negligible since the cylinder is far from the domain boundary, where the zero-gradient boundary conditions are imposed. The Eulerian grid is uniform in all directions, and the Lagrangian points used by the IBM method are equally distributed on the cylinder surface. The grid size is $dx=dy=D/64$. The simulations have been performed on the HPE Cray EX supercomputer Dardel at the PDC Center for High-Performance Computing at KTH, and each wake simulation took 98.000 core hours on average.The time step was kept constant at $\Delta t=0.0015$.

\subsection{Newtonian flow validation}\label{section:Newtonian_verification}
\begin{figure} 
\begin{subfigure}{0.5\textwidth}
\centering
\includegraphics[width=1.0\linewidth,height=3.0cm]{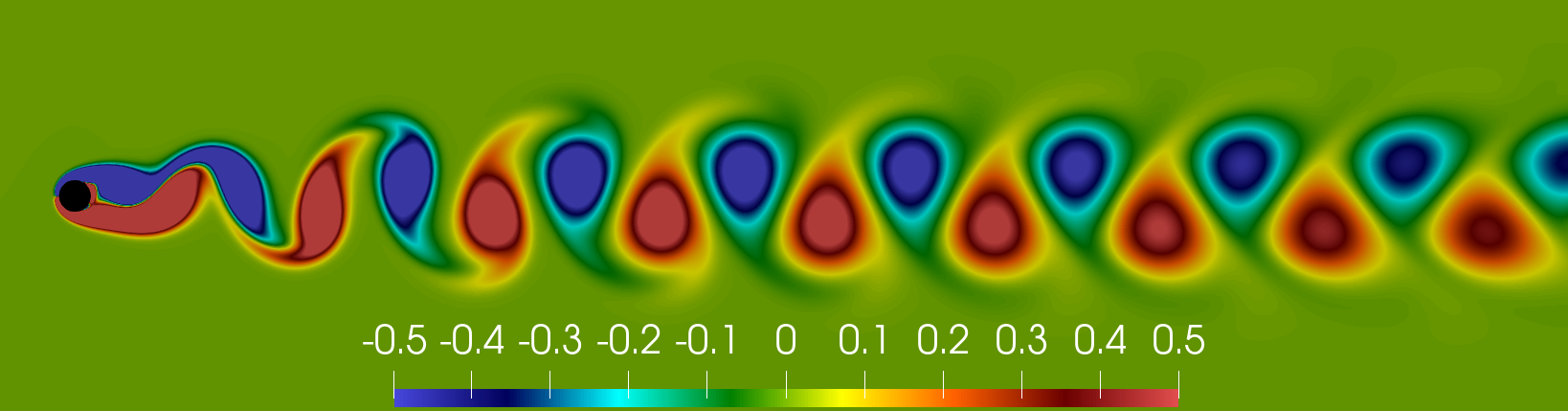}
\caption{Vorticity $\zeta_{z}$}
\label{fig:subim2_1}
\end{subfigure}
\begin{subfigure}{0.5\textwidth}
\centering
\includegraphics[width=1.0\linewidth,height=3.0cm]{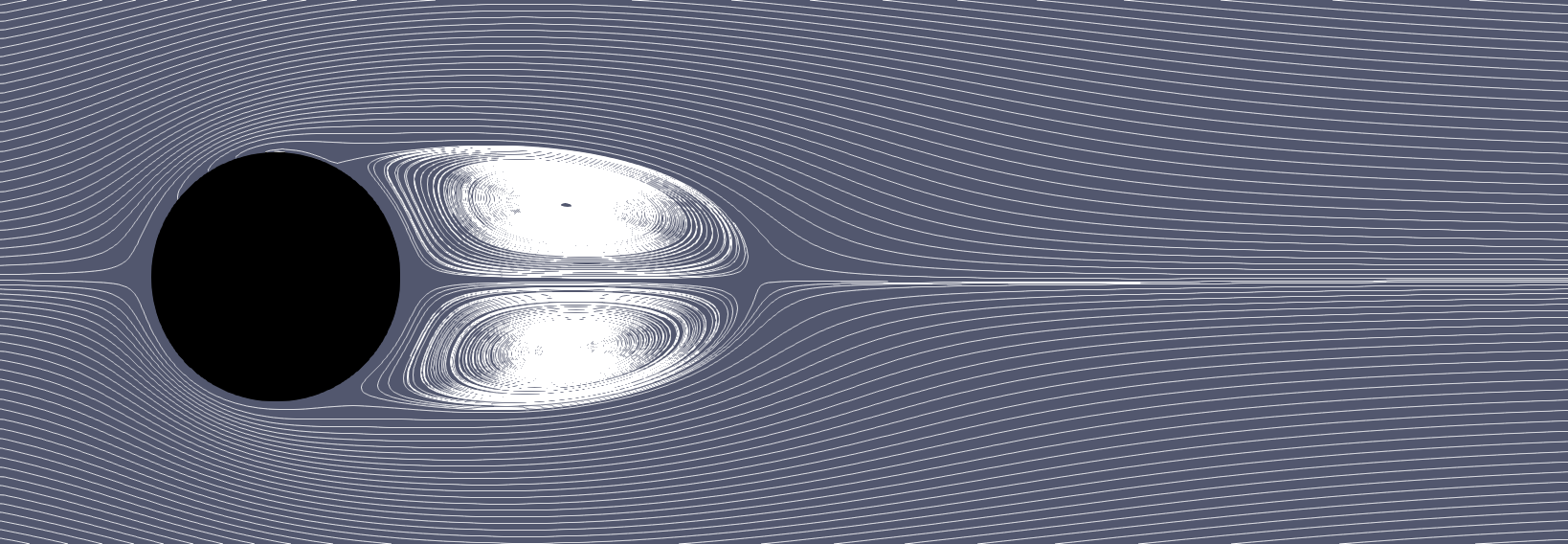}
\caption{Streamlines}
\label{fig:subim2_2}
\end{subfigure}
\caption{The vorticity $\zeta_{z}$ (a) and the streamlines (b) of a Newtonian fluid flow past a cylinder, at $Re=100$}
\label{fig:image2}
\end{figure}

Here, the numerical setup is validated against data of two-dimensional steady and unsteady flows of Newtonian fluid at $Re=30-100$ available in the literature. For $Re = 30$, two attached, symmetric, closed eddies which create a steady recirculating region are observed just behind the cylinder, as discussed in \cite{Williamson1996,Posdziech2007,Constant2017}, not presented here for brevity. At $Re = 50$, and subsequent $Re=100$ cases, the instability grows, and a periodic flow fluctuation appears. \Figref{fig:subim2_1} shows the evolution of the vorticity contours $\zeta_{z}$ at $Re = 100$ and the well-known \vonK vortex street.
As shown, the vortex shedding creates an upper row  of cyclonic (negative, blue in the figure) and a lower row of anticyclonic (positive, red in the figure) vortices parallel to the cylinder axis, and a recirculation region attached to the rear of the cylinder. The vortices travel downstream and gradually change to tear-shaped vortices. \Figref{fig:subim2_2} shows the time-averaged streamlines of the flow, and the length of recirculation  bubble is $L_{RB}=1.42$, as reported in \Tabref{tab:1}.

The comparison of the steady and unsteady flow features for $Re = 30,\, 50$, and $100$ with previous studies 
shows a very good agreement in predicting the drag coefficient $C_D$, the root mean square (rms) of the lift coefficient $C_{L}$, the Strouhal number $St$, the time-averaged length of recirculation region, and the separation angle, see \Tabref{tab:1}. 
It is worth mentioning that to calculate the $St$ number, we extract the time history of the normal velocity downstream of the cylinder. The primary flow frequency $f_{v}$ is obtained from the peak frequency of Fast Fourier Transform (FFT), and then used to calculate the $St$ number. 
Finally in the table, we also report the ratio between the rms of the drag to lift coefficients $r$. This changes from periodic to chaotic when the magnitude of $r$ decreases to $r\sim O(1)$. Increasing $Re$ decreases the mean $C_{D}$, $\Theta_{s}$ and $r$ ratio, while  $(C_{L_{rms}})$ and $St$ increase. These observations were also reported in \cite{Williamson1996,Mossaz2010,Qu2013,Jiang2020}.

 \begin{table} 
   \begin{center}
 \def~{\hphantom{0}}
   \begin{tabular}{lccccccc}
   $ Study $                                   &$C_{D}$&$C_{L_{rms}}$&$St$ &$\Theta_{s}$&$L_{RB}$&   r  \\[2pt]\hline  
   $Re=$ 30                                    &       &             &     &              &                  &      \\     \hline
   Tritton \cite{Tritton1959}                  &1.74   &   ~~-~~     & -   &              &                  &      \\
   Coutanceau and Bouard\cite{Coutanceau1977}  &~-~    &   ~~-~~     & -   &    130.0     &     1.55         &      \\
   Pinelli et al. \cite{Pinelli2010}           &1.80   &   ~~-~~     & -   &    131.9     &     1.70         &      \\
   Constant et al.\cite{Constant2017}          &1.77   &   ~~-~~     & -   &    131.6     &     1.64         &      \\     
   (Present 32$\times$16)                      &1.79   &   0.        & 0.  &    131.8     &     1.73         &      \\     
   (Present 64$\times$32)                      &1.74   &   0.        & 0.  &    131.4     &     1.65         &      \\     \hline   
   $Re=$ 50                                    &~~~~~  &   ~~~~~     &     &              &                  &      \\     \hline
   Oliveira \cite{Oliveira2001}                &1.476  &   ~~-~~     &0.126&     ~-~~     &     2.39         &      \\    
   Sivakumar et al. \cite{Sivakumar2006}       &1.41~  &   0.034     &0.122&    ~-~       &                  &      \\        
   Mossaz et al. \cite{Mossaz2010}             &1.418  &   ~~-~~     &0.122&    ~-~       &                  &      \\        
   Qu et al. \cite{Qu2013}                     &1.397  &   0.040     &0.124&    124.0     &     2.46         &327   \\      
   (Present 32$\times$16)                      &1.506  &   0.053     &0.127&    123.4     &     2.50         &176.0 \\     
   (Present 64$\times$32)                      &1.456  &   0.043     &0.125&    124.1     &     2.62         &280.5 \\     \hline
   $Re=$ 100                                   &~~~~~  &   ~~~~~     &     &              &                  &      \\     \hline
   Williamson \cite{Williamson1996}            &~-~~~  &   ~~-~~     &0.165&     ~-~      &                  &      \\        
   Kravchenko et al.\cite{Kravchenko1997}      &1.314  &   0.222     &0.164&    117.4     &     1.45         &      \\
   Oliveira \cite{Oliveira2001}                &1.370  &   ~~-~~     &0.167&    ~~-~~     &     1.40         &      \\    
   Sivakumar et al. \cite{Sivakumar2006}       &1.325  &   0.229     &0.164&    ~~-~~     &                  &      \\        
   Posdziech and Grundmann\cite{Posdziech2007} &1.325  &   0.228     &0.164&    ~~-~~     &                  &      \\    
   Mossaz et al. \cite{Mossaz2010}             &1.328  &   ~~-~~     &0.164&    ~~-~~     &                  &      \\
   Qu et al. \cite{Qu2013}                     &1.319  &   0.225     &0.165&    118.0     &     1.41         &36.0  \\
   Constant et al. \cite{Constant2017}         &1.37~  &   ~~-~~     &0.165&    118.9     &                  &      \\
   Peng et al. \cite{Peng2021}                 &1.361  &   0.235     &0.165&    ~~-~~     &                  &      \\
   (Present 32$\times$16)                      &1.400  &   0.254     &0.168&    118.0     &     1.40         &34.0  \\
   (Present 64$\times$32)                      &1.359  &   0.236     &0.164&    117.5     &     1.41         &34.7  \\     \hline 

   \end{tabular}
   \caption{Cylinder flow characteristics (the drag coefficient $C_D$, the rms of lift coefficient $C_{L_{rms}}$, Strouhal number $St$, separation angle $\Theta_{s}$, the recirculation bubble length $L_{RB}$, and the ratio of fluctuations of drag and lift coefficients $r=\frac{(C_{L_{rms}})}{(C_{D_{rms}})}$ at different Reynolds number $(Re=30,50,100)$ for Newtonian fluid}
   \label{tab:1}
   \end{center}
 \end{table}

\begin{figure} 
\begin{subfigure}{0.5\textwidth}
\centering
\includegraphics[width=0.8\linewidth, height=5cm]{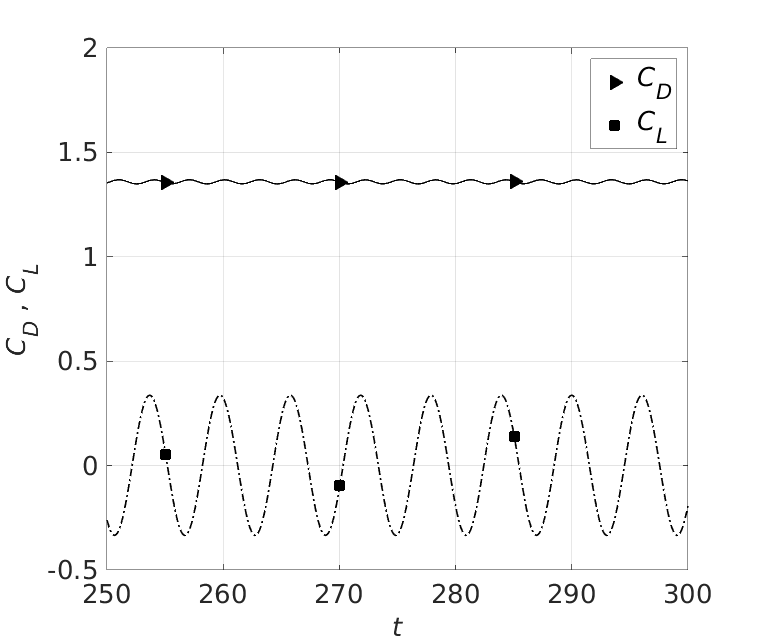}
\caption{\label{Re100N_CDCL} $C_D$ and $C_L$  }
\label{fig:subim3_1}
\end{subfigure}
\begin{subfigure}{0.5\textwidth}
\centering
\includegraphics[width=0.8\linewidth, height=5cm]{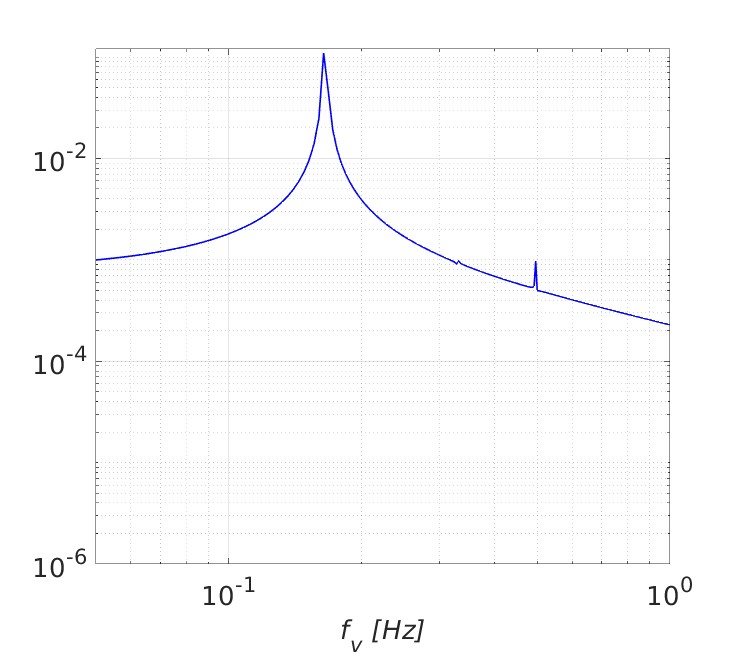}
\caption{\label{Re100N_FFT} $f_{v}$}
\label{fig:subim3_2}
\end{subfigure}
\caption{The drag and lift coefficients variation (a) and frequency $f_{v}$ (b) of a Newtonian fluid flow past a cylinder for $Re=100$.}
\label{fig:image3}
\end{figure}

\Figref{fig:subim3_1} displays the time evolution of $C_D$ and $C_L$ for $Re=100$. As seen, these coefficients clearly show a sinusoidal variation. We observe that the mean drag coefficient decreases by increasing $Re$. However, the amplitude of the lift and drag coefficients' ﬂuctuations increases with increasing inertia. These observations are consistent with the previous studies in \Tabref{tab:1}. Finally, \Figref{fig:subim3_1} shows the frequency of the velocity fluctuations. As expected at $Re=100$ for a Newtonian fluid, there is only one primary frequency.

As Constant \etal \cite{Constant2017} mentioned, the grid size $D/50$ is sufficient for a numerical simulation of Newtonian fluid past a cylinder. Here, we also reach the same conclusion. However, since the fluid of interest is EVP, a finer mesh with a grid $D/64$ size is used in the present study. It is also observed that a domain size with $32D \times 16D$ is too small and may negatively affect the results, as shown in \Tabref{tab:1}. Therefore a larger domain with $64D \times 32D$ is used in the present work.

\subsection{Non-Newtonian validation}\label{section:nonNewtonian_fluids}
\begin{figure} 
\begin{subfigure}{0.5\textwidth}
\centering
\includegraphics[width=0.9\linewidth,height=5.5cm]{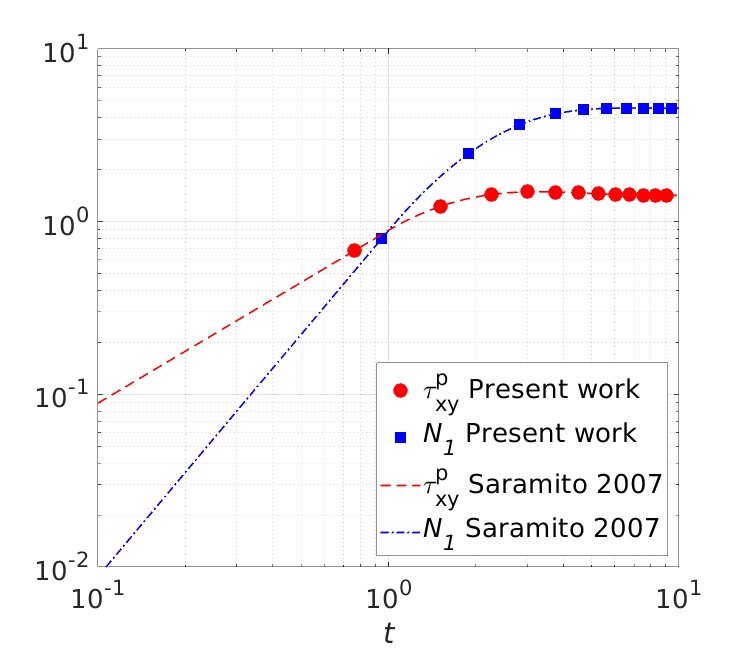}
\caption{Saramito Oldroyd-B Model $n=1.0$}
\label{fig:subim4_1}
\end{subfigure}
\begin{subfigure}{0.5\textwidth}
\centering
\includegraphics[width=0.9\linewidth,height=5.5cm]{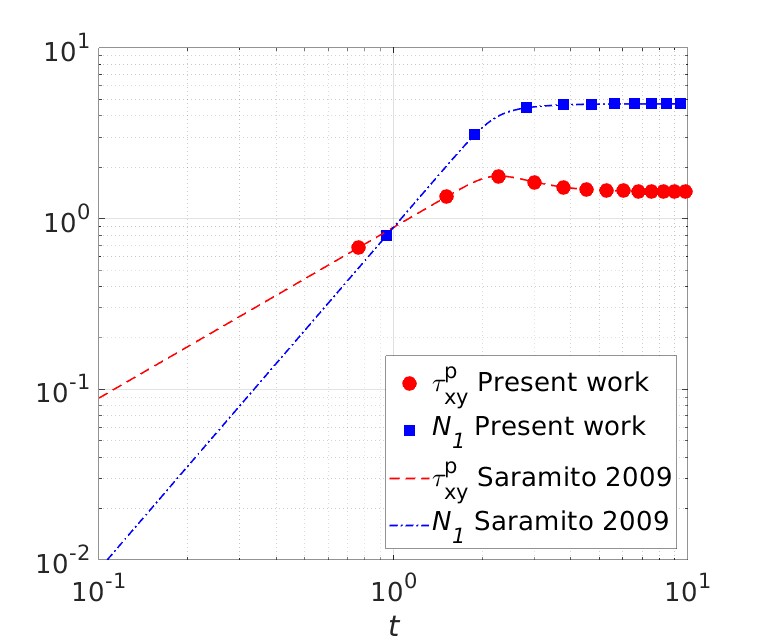}
\caption{Saramito Herschel-Bulkley Model $n=0.2$}
\label{fig:subim4_2}
\end{subfigure}
\caption{Simple shear flow problem, the comparison of present work with a semi-analytical solution of Saramito Oldroyd-B Model \cite{Saramito2007} (a) and Saramito Herschel-Bulkley Model \cite{Saramito2009} (b) for $Re=0.1, Wi=1.0,\beta_{s}=1./9.$ and $Bn=1.0$.}
\label{fig:image4}
\end{figure}

The in-house IBM code is verified against several canonical problems as reported in \cite{Izbassarov2018} and has been successfully used in previous studies of EVP fluids with particles \cite{Chaparian20201}, in porous media \cite{Chaparian20202,Chaparian2021} and turbulent channel flow \cite{Izbassarov2021}. For completeness, we show a validation of the EVP model against semi-analytical solutions under simple shear flow in \Figref{fig:image4}. The simulations are performed at $Re=0.1,\, Wi=1,\,\beta_{s}=1./9.$ and $Bn=1$, and two shear-thinning indices: $n=0.2$, and $n=1$ (for which the model reduces to the original Saramito model without shear-thinning \cite{Saramito2007}). As shown in \Figref{fig:image4} the results of both simulations overlap perfectly with semi-analytical solutions.

\section{Results and Discussion}\label{section:Results_Discussion}
In this section, we first present the results for a viscoelastic fluid and then discuss the combined effects of yield stress and elasticity on the flow around a cylinder. In all cases, the Reynolds number is kept constant at $Re=100$, in which the flow is unsteady and has vortex shedding in a Newtonian fluid, so we can study the effect of elastic and plastic forces on the flow pattern.

\subsection{Viscoelastic cylinder}\label{section:Oldroyd-B_Model}

The main features of the viscoelastic cylinder flow are reported in 
\Tabref{tab:2} for increasing elasticity.
The viscoelastic fluid is described by the Saramito model with zero yield stress, in which case the model reduces to the Oldroyd-B viscoelastic model. 

The data in the table are visualised in \Figref{fig:image5}. The average $C_D$ for the viscoelastic fluid cases is slightly higher than the average $C_D$ for the Newtonian fluid flow at $Re=100$. However, $C_D$ does not increase further with $Wi$ but stays practically constant, from $C_D=1.431$ at $Wi=1$, to $C_D=1.393$ at $Wi=4$, as shown in panel \ref{fig:subim5_1}. 
\Figref{fig:subim5_2} shows that the root mean square (rms) of the lift coefficient $C_{L_{rms}}$ on the other hand decreases significantly with $Wi$, which indicates a smaller fluctuation amplitude. The Strouhal number has a tiny decrease at $Wi=1, 2$ compared to the Newtonian case ($St=0.164$). However, at a larger $Wi$ number, $Wi=4$, instead of periodic and sinusoidal behaviour, we find an irregular behaviour of the instantaneous lift and drag coefficients, and the flow becomes more chaotic in time. We therefore do not report $St$ for this flow case. 
A similar finding was also reported by Peng \etal \cite{Peng2021}. The separation angle $\Theta_{s}$ reduces for $Wi=1$, but increases thereafter with increasing $Wi$ up to $Wi=4$. The increasing elasticity also increases $L_{RB}$, see \Figref{fig:subim5_3}. This observation is consistent with former experimental and numerical studies \cite{Oliveira2001,Richter2010,Peng2021}.

Increasing polymer concentration to $\beta_s=0.5$ results in a modest increase of the mean of drag coefficient to $C_D=1.521$ and a sharp decrease of the lift coefficient to $C_{L_{rms}}=0.076$, as shown in \Figref{fig:subim5_1} and \Figref{fig:subim5_2}. It is observed that the amplitude of $C_L$ decreases by increasing elasticity \ie increasing $Wi$ or decreasing $\beta_{s}$, which shows the instability is suppressed in the presence of the polymer.

\begin{table} 
   \begin{center}
 \def~{\hphantom{0}}
   \begin{tabular}{lccccccc}
   $ Case $&$C_{D}$ &$C_{L_{rms}}$ & $St$ &$\Theta_{s}$ &$L_{RB}$& r \\[2pt]\hline
   $\beta_{s}=0.9$&        &              &      &              &                   &       \\
   $Wi=1.0$&1.431~  &   0.188~     &0.156~&   ~114.1~    &     1.67~         &35.5~  \\     
   $Wi=2.0$&1.405~  &   0.069~     &0.150~&   ~115.1~    &     2.42~         &~6.0~  \\     
   $Wi=4.0$&1.393~  &   0.033~     &~-~   &   ~116.2~    &     3.07~         &~1.4~  \\\hdashline 
   $\beta_{s}=0.5$&        &              &      &              &                   &       \\[2pt]     
   $Wi=1.0$&1.521~  &   0.076~     &0.14~ &   ~111.8~    &     2.52~         &~5.2~  \\\hline     
   \end{tabular}
   \caption{Flow characteristics (the drag coefficient $C_D$, the rms of lift coefficient $C_{L_{rms}}$, Strouhal number $St$, separation angle $\Theta_{s}$, the recirculation bubble length $L_{RB}$, and the ratio of fluctuations of drag and lift coefficients $r=(\frac{C_{L_{rms}}}{C_{D_{rms}}})$ for the viscoelastic flow past a circular cylinder at $Re=100$, $Wi=1\to 4$, $n=1$ and $Bn=0$.}
   \label{tab:2}
   \end{center}
 \end{table}

\Pab{
\begin{figure} 
\begin{subfigure}{0.32\textwidth}
\includegraphics[width=0.95\linewidth, height=4cm]{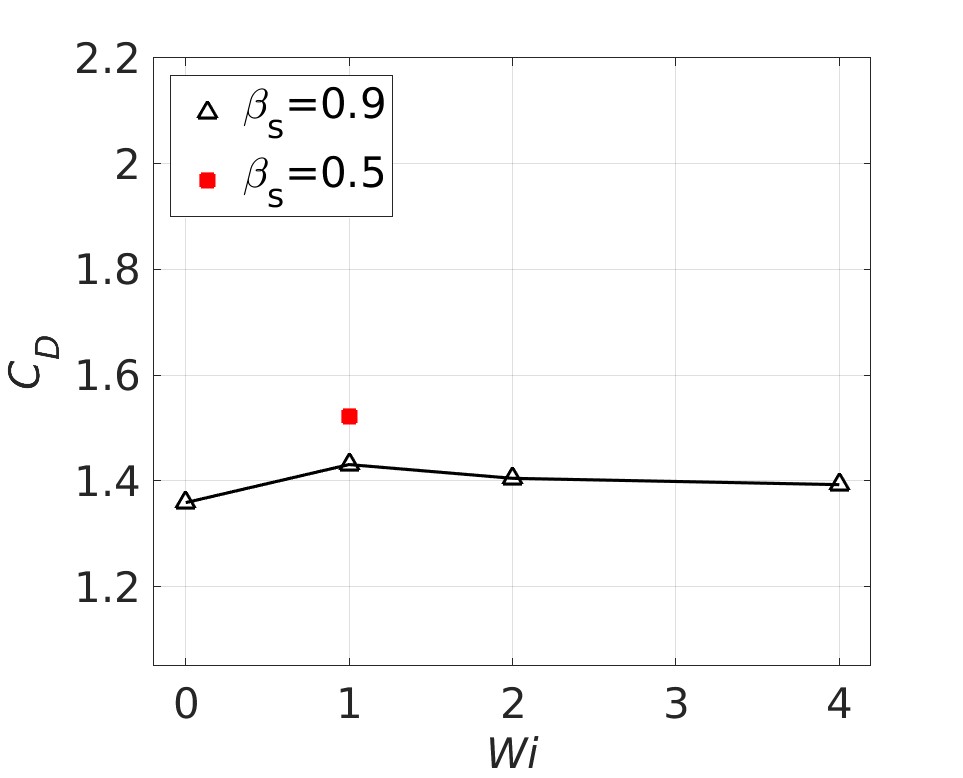}
\caption{\label{VE_Cd} $C_{D}$ }
\label{fig:subim5_1}
\end{subfigure}
\begin{subfigure}{0.34\textwidth}
\includegraphics[width=1.0\linewidth, height=4cm]{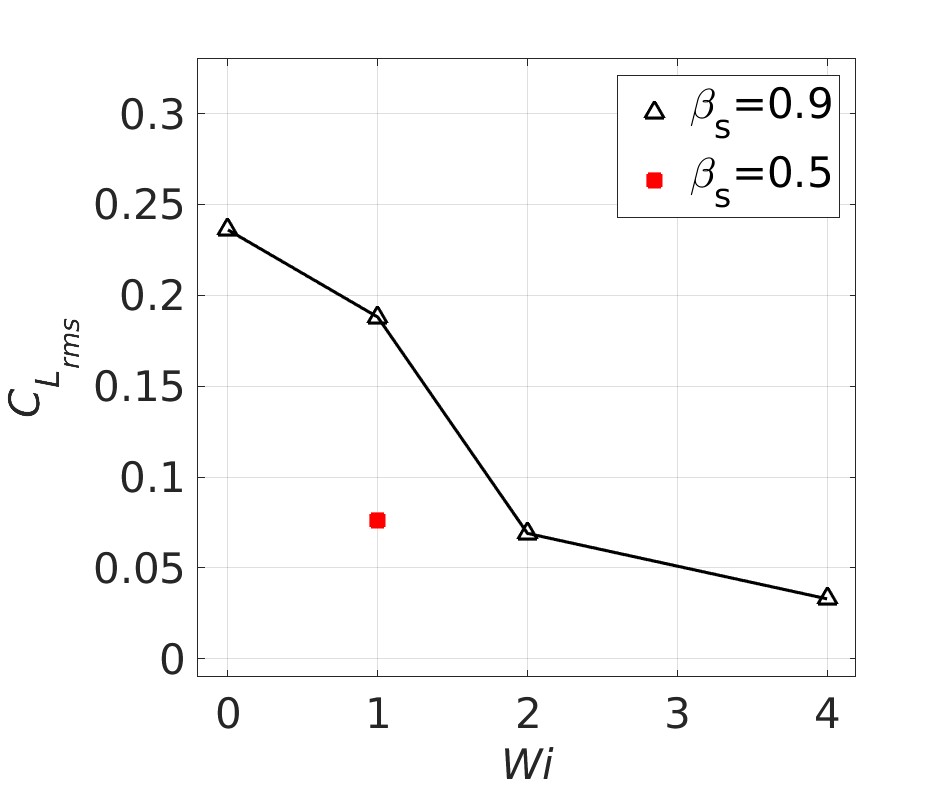}
\caption{\label{VE_CL} $C_{L_{rms}}$ }
\label{fig:subim5_2}
\end{subfigure}
\begin{subfigure}{0.32\textwidth}
\centering
\includegraphics[width=0.95\linewidth, height=4.0cm]{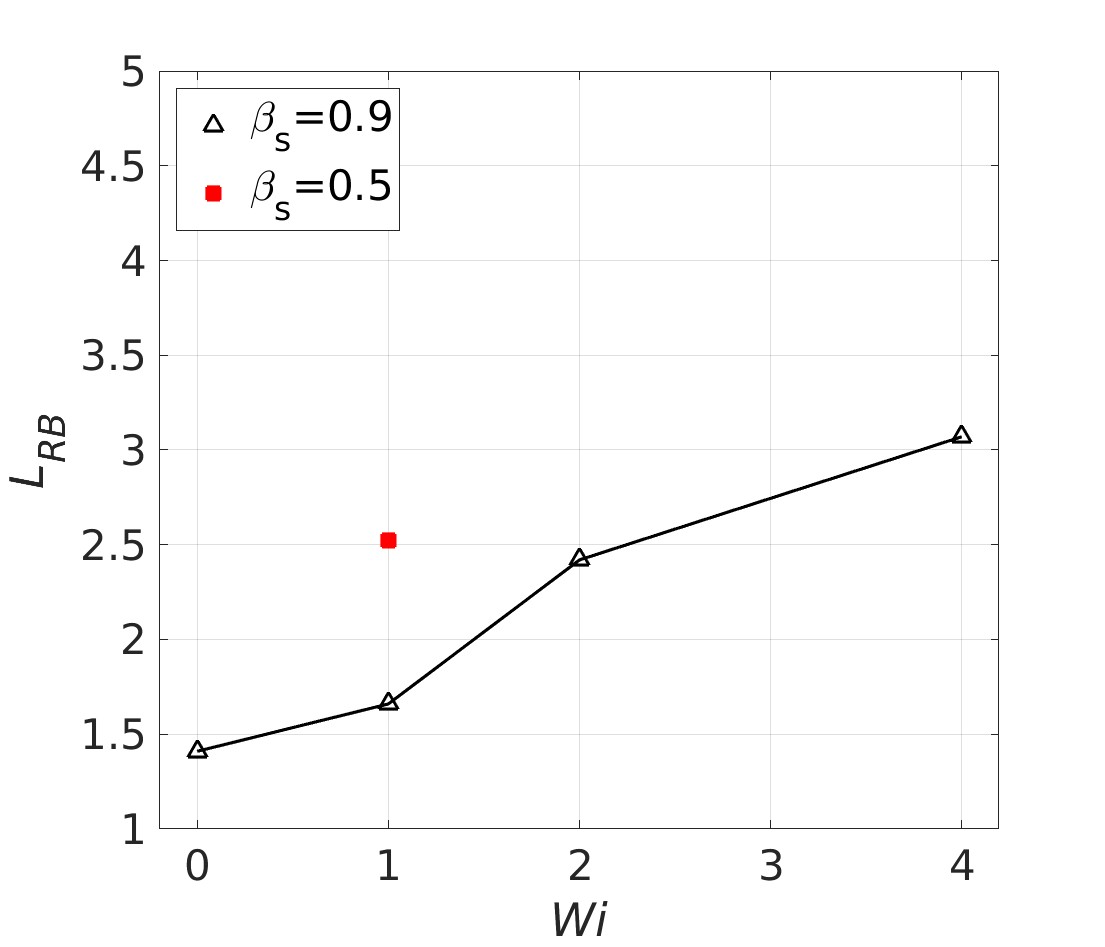}
\caption{\label{VE_Recirculation} $L_{RB}$}
\label{fig:subim5_3}
\end{subfigure}
\caption{(a) The mean of drag coefficient; (b) the root mean square of the lift coefficient and (c) the length of recirculation bubble ($L_{RB}$)for the flow past a circular cylinder of a viscoelastic fluid with $Re=100$,  $Wi=1.0-4.0, \beta_{s}=0.5$ (close marker) $\beta_{s}=0.9$ (open markers), $n=1.0$, $Bn=0.0$}
\label{fig:image5}
\end{figure}
}

\begin{figure} 
\begin{subfigure}{0.6\textwidth}
\centering
\includegraphics[width=1.0\linewidth, height=3.0cm]{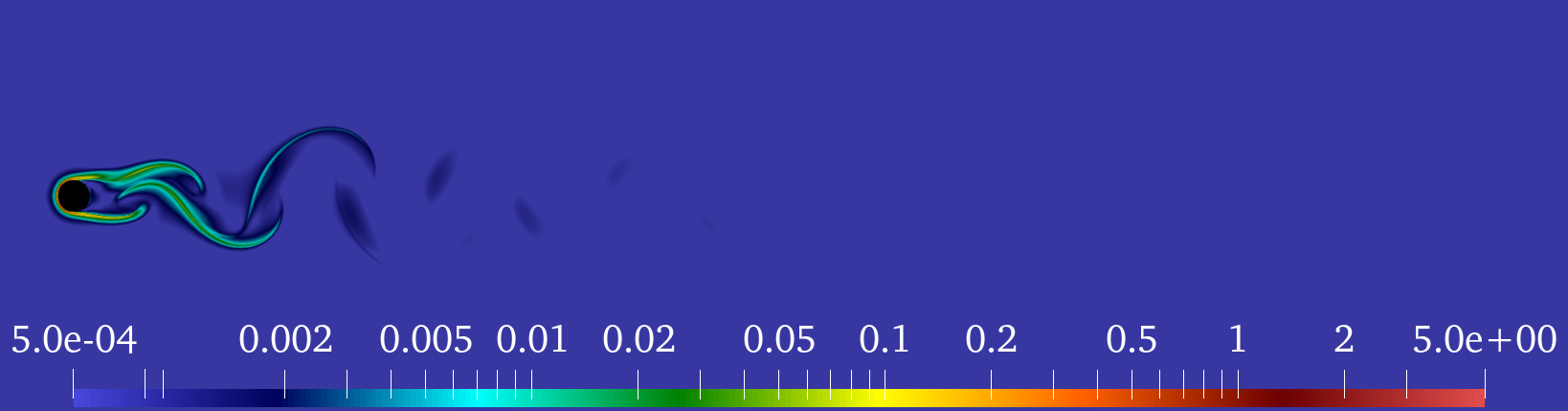}
\caption{$ Wi=1.0$ }
\label{fig:subim6_1}
\end{subfigure}
\begin{subfigure}{0.4\textwidth}
\centering
\includegraphics[width=1.0\linewidth, height=3.0cm]{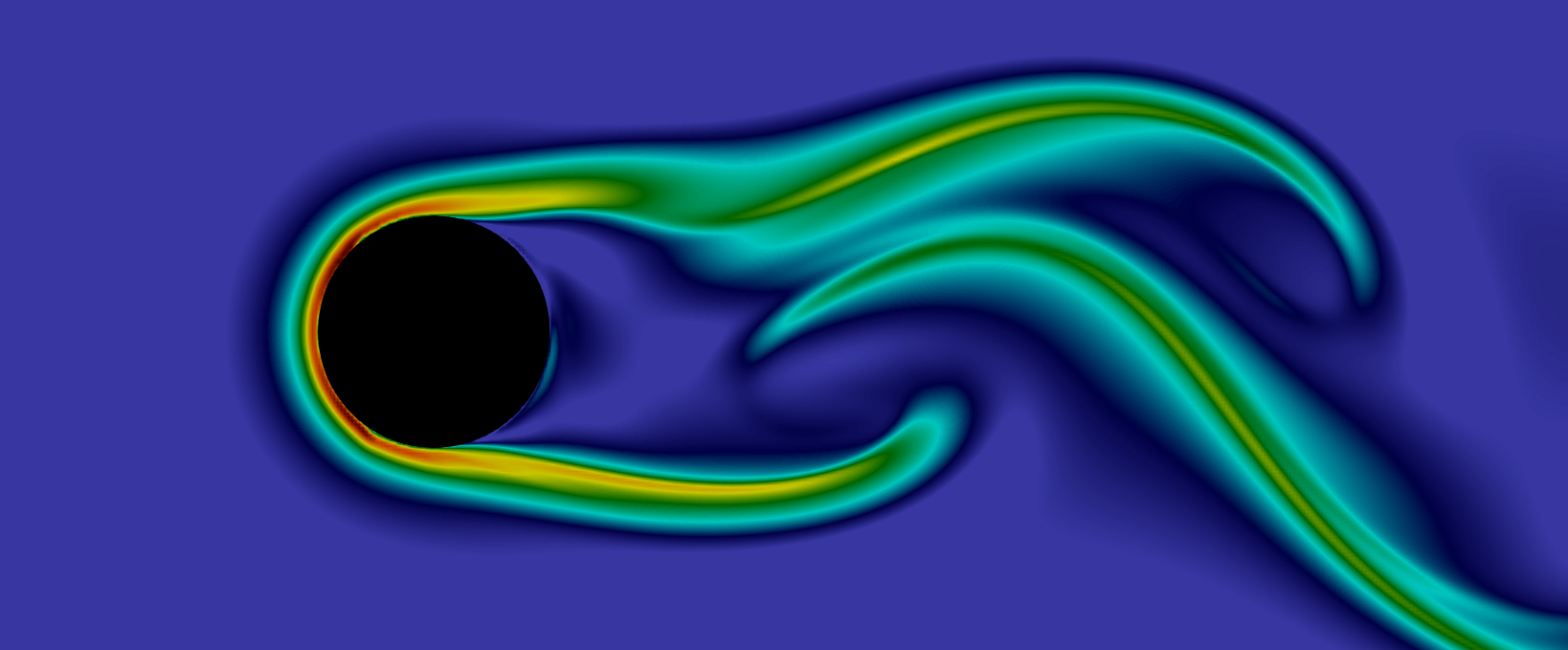}
\caption{The zoom of \Figref{fig:subim6_1}}
\label{fig:subim6_2}
\end{subfigure}
\begin{subfigure}{0.6\textwidth}
\centering
\includegraphics[width=1.0\linewidth, height=3.0cm]{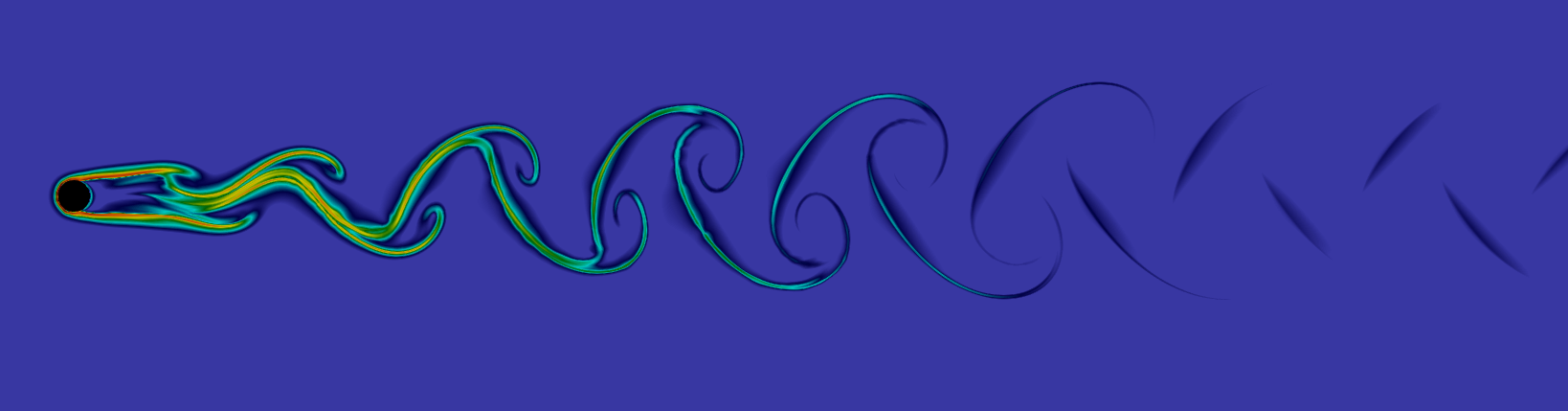}
\caption{$ Wi=4.0$ }
\label{fig:subim6_3}
\end{subfigure}
\begin{subfigure}{0.4\textwidth}
\centering
\includegraphics[width=1.0\linewidth, height=3.0cm]{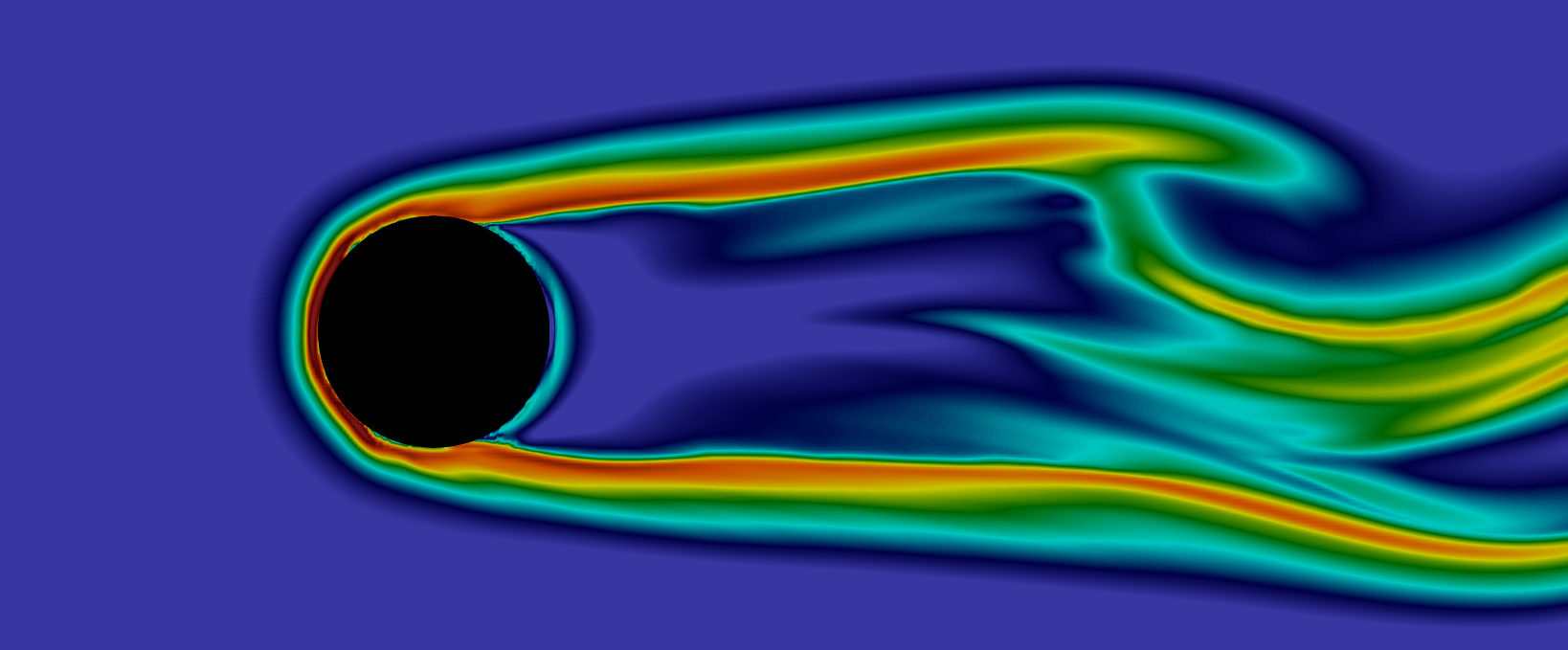}
\caption{The zoom of \Figref{fig:subim6_3}}
\label{fig:subim6_4}
\end{subfigure}
\caption{Contours of the trace of the polymer stress tensor, $\trace{\btau^p}=\tau_{xx}^{p}+\tau_{yy}^{p}+\tau_{zz}^{p}$, for a viscoelastic fluid at $Wi=$1.0 and 4.0, $\beta_{s}=0.9$, $n=1.0$, and $Bn=0.0$, at $Re=100$. The color scale is the same in all plots.}
\label{fig:image6}
\end{figure}

\Figref{fig:image6} shows the contour of the trace of polymer stress tensor $\trace{\btau^p}=\tau_{xx}^{p}+\tau_{yy}^{p}+\tau_{zz}^{p}$ at $Wi=1.0,4.0$ for a dilute solution with $\beta_{s}=0.9$, where the trace of the tensor is proportional to the elongation of the polymer chains. On the left column of \Figref{fig:image6}, the entire domain is depicted, while on the right, we focus on the region around the cylinder to provide more details. 
As can be seen in \Figref{fig:image6}, by increasing $Wi$ from $Wi=1$ to $Wi=4$,
the maximum extension closer to the cylinder surface increases (shown by the increase of the global maximum of $\trace{\btau^p}$)
and also regions with higher polymer extensions can be found further downstream of the cylinder surface.
The same behaviour is observed for $\trace{\btau^p}$ when increasing polymer concentration ($\beta_s=0.5$), not shown for the sake of brevity. For all cases, the trace of the polymer stress tensor gradually decays with the distance downstream of the cylinder, from $\trace{\btau^p} \approx 5$, close to the cylinder, to $\trace{\btau^p} \approx 5\times10^{-4}$ at a large distance downstream of the cylinder. 

\begin{figure} 
\begin{subfigure}{0.6\textwidth}
\centering
\includegraphics[width=1.0\linewidth, height=3.0cm]{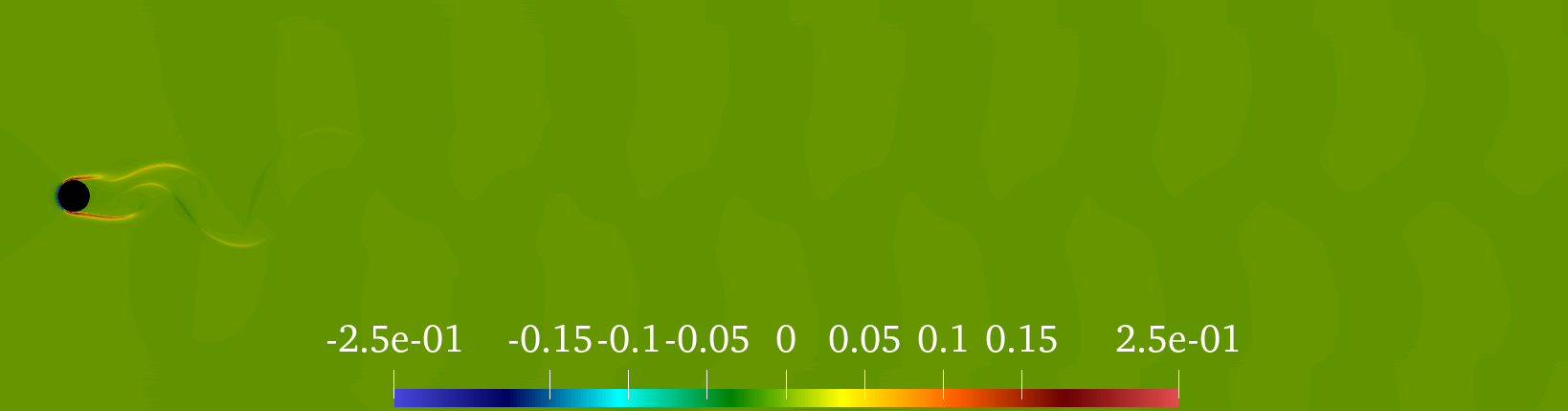}
\caption{$ Wi=1.0$ }
\label{fig:subim7_1}
\end{subfigure}
\begin{subfigure}{0.4\textwidth}
\centering
\includegraphics[width=1.0\linewidth, height=3.0cm]{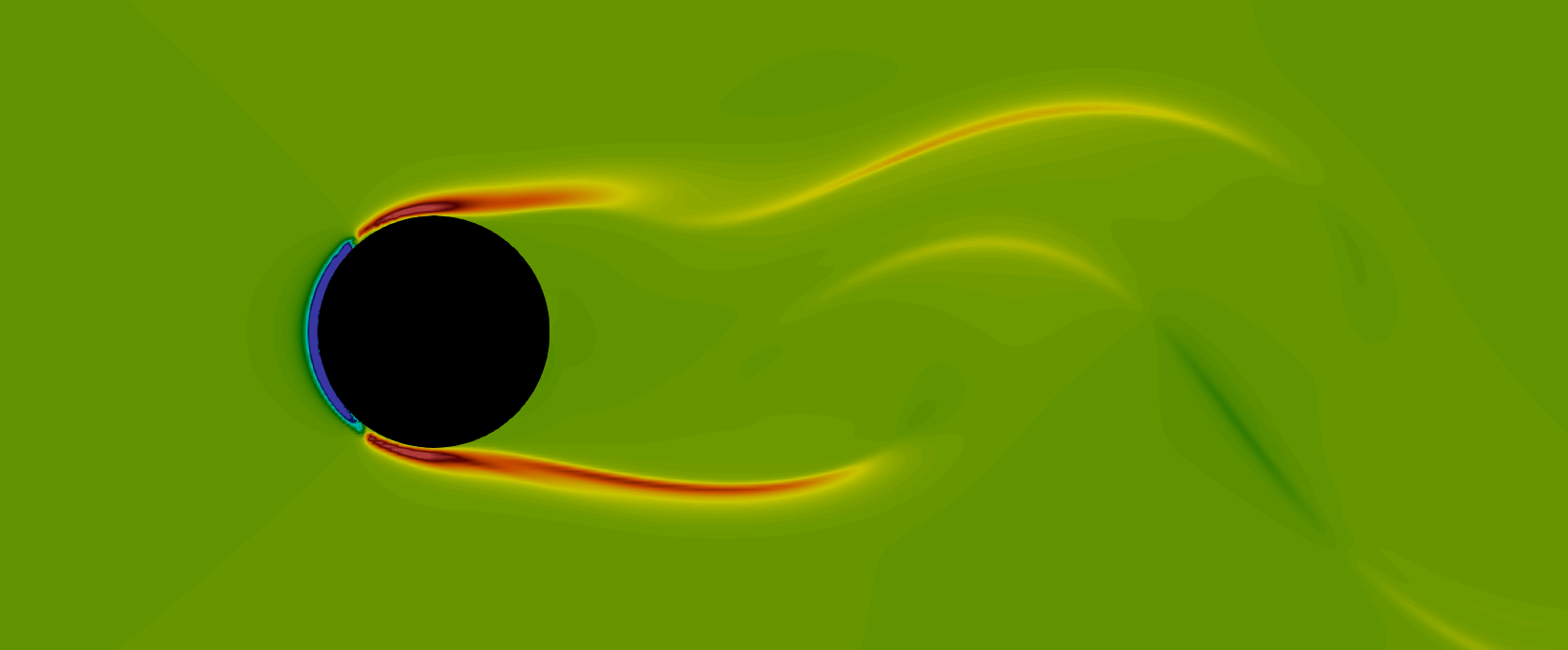}
\caption{The zoom of \Figref{fig:subim7_1}}
\label{fig:subim7_2}
\end{subfigure}
\begin{subfigure}{0.6\textwidth}
\centering
\includegraphics[width=1.0\linewidth, height=3.0cm]{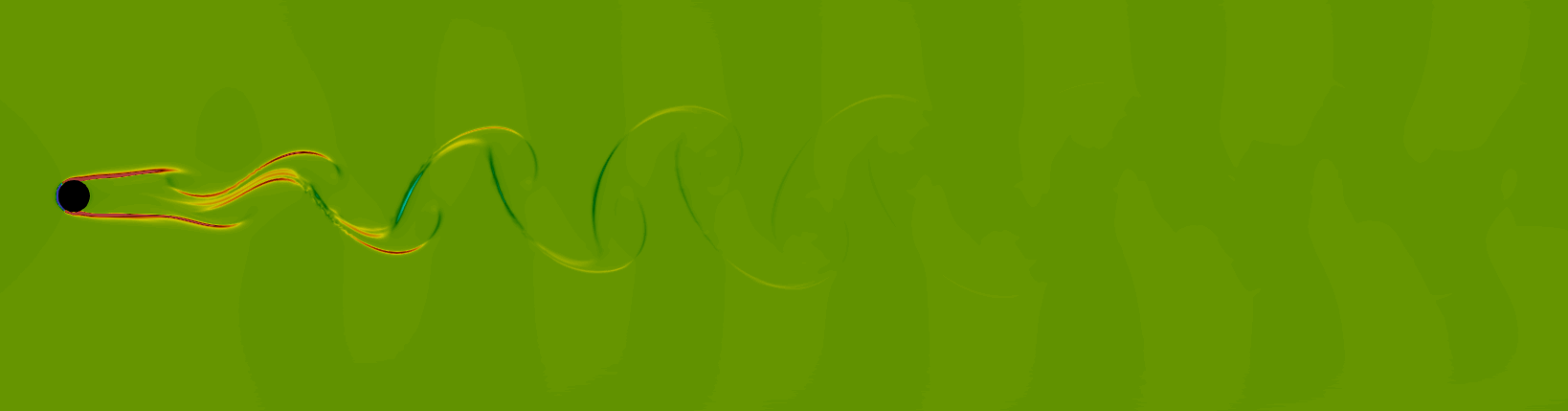}
\caption{$ Wi=4.0$ }
\label{fig:subim7_3}
\end{subfigure}
\begin{subfigure}{0.4\textwidth}
\centering
\includegraphics[width=1.0\linewidth, height=3.0cm]{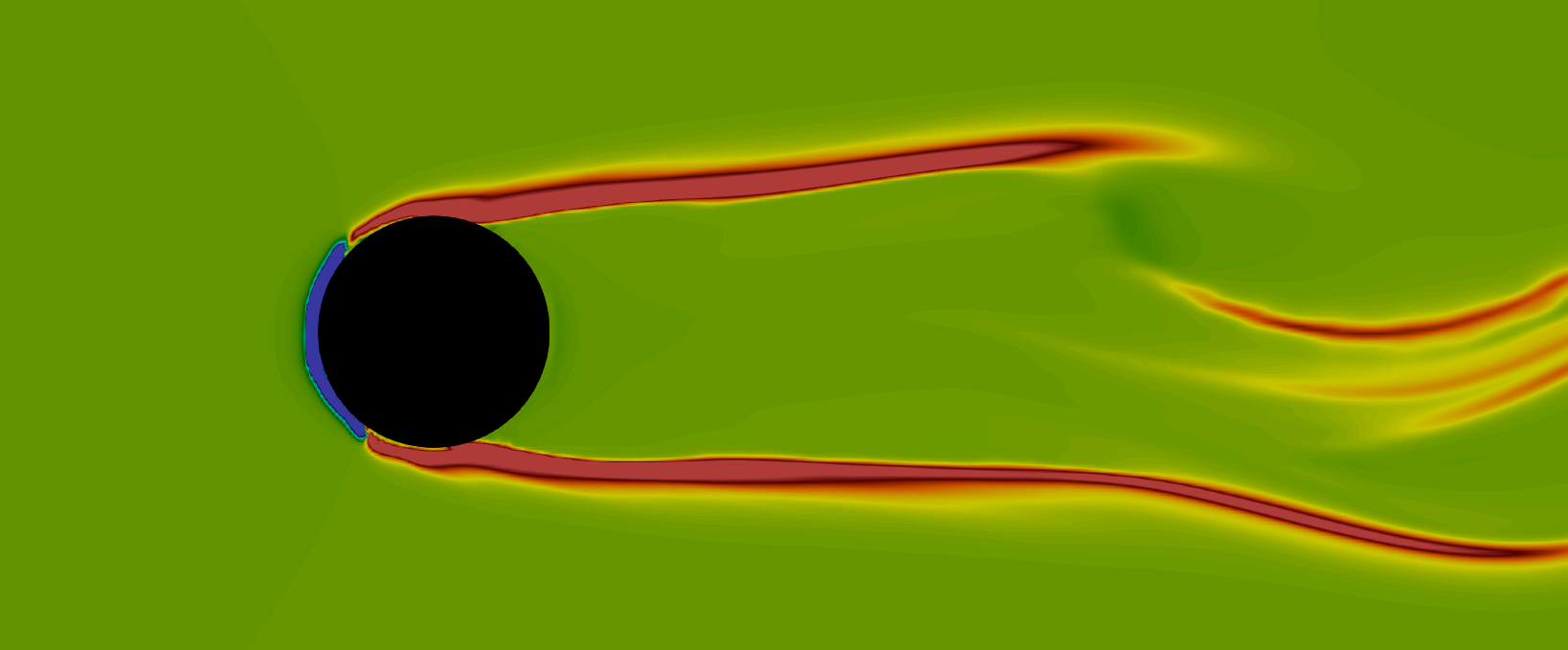}
\caption{The zoom of \Figref{fig:subim7_3}}
\label{fig:subim7_4}
\end{subfigure}
\caption{Contours of the first normal stress difference contour $N_{1}=\tau_{xx}^{p}-\tau_{yy}^{p}$ for the viscoelastic flow past a circular cylinder with $Wi=1.0$ and 4. (top and bottom), and $\beta_{s}=0.9$, $n=1.0$, and $Bn=0.0$ at $Re=100$. The color scale of the two contour plots is the same.}
\label{fig:image7}
\end{figure}

The appearance of a normal stress is not only one of the most interesting effects of a complex rheology but also one of the primary non-linear effects. Contours of the first normal stress difference $N_{1}=\tau_{xx}^{p}-\tau_{yy}^{p}$ are
 shown in \Figref{fig:image7}. Here we note that 
increasing elasticity increases the magnitude of $N_{1}$. This quantity is positive over the whole domain except close to the cylinder upstream stagnation point, where $N_{1}\leq 0$. 
As a consequence,  the magnitude of normal stress $\tau_{xx}^{p}$ is larger than $\tau_{yy}^{p}$ over the whole domain except at the upstream stagnation point. 
This is attributed to the change in the flow direction from predominantly streamwise to cross-stream, to align to the cylinder surface. 
We also notice that the maximum $\tau_{yy}^{p}$ stress always occurs at the upstream stagnation point and it intensifies when increasing the elasticity, \ie increasing $Wi$ or decreasing $\beta_{s}$. 

\begin{figure} 
\begin{subfigure}{0.5\textwidth}
\centering
\includegraphics[width=1.0\linewidth, height=3.0cm]{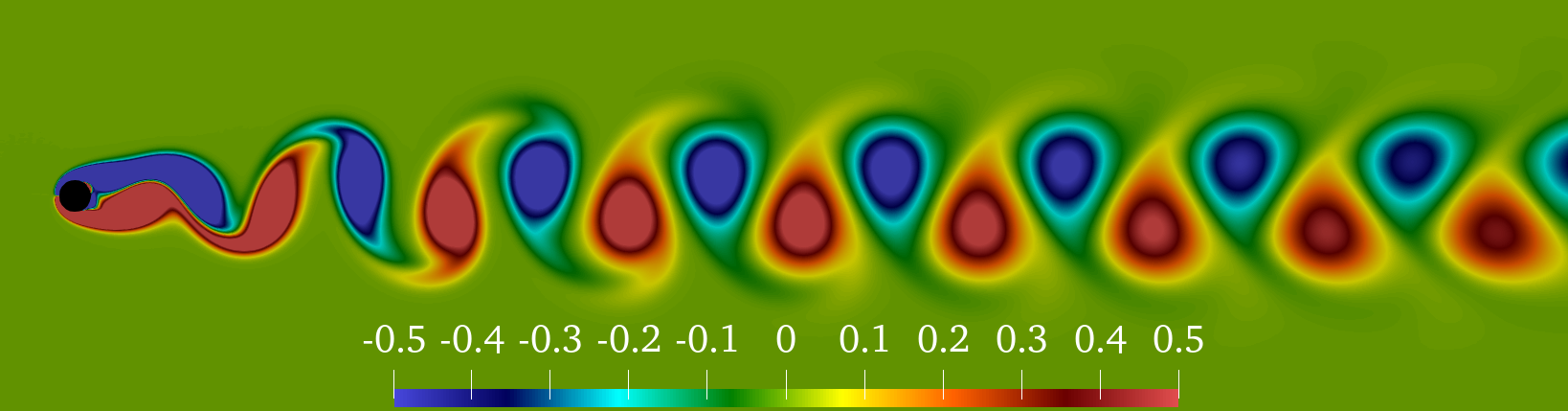}
\caption{Vorticity $\zeta_{z}$ $(Wi=1.0)$ }
\label{fig:subim8_1}
\end{subfigure}
\begin{subfigure}{0.5\textwidth}
\centering
\includegraphics[width=1.0\linewidth, height=3.0cm]{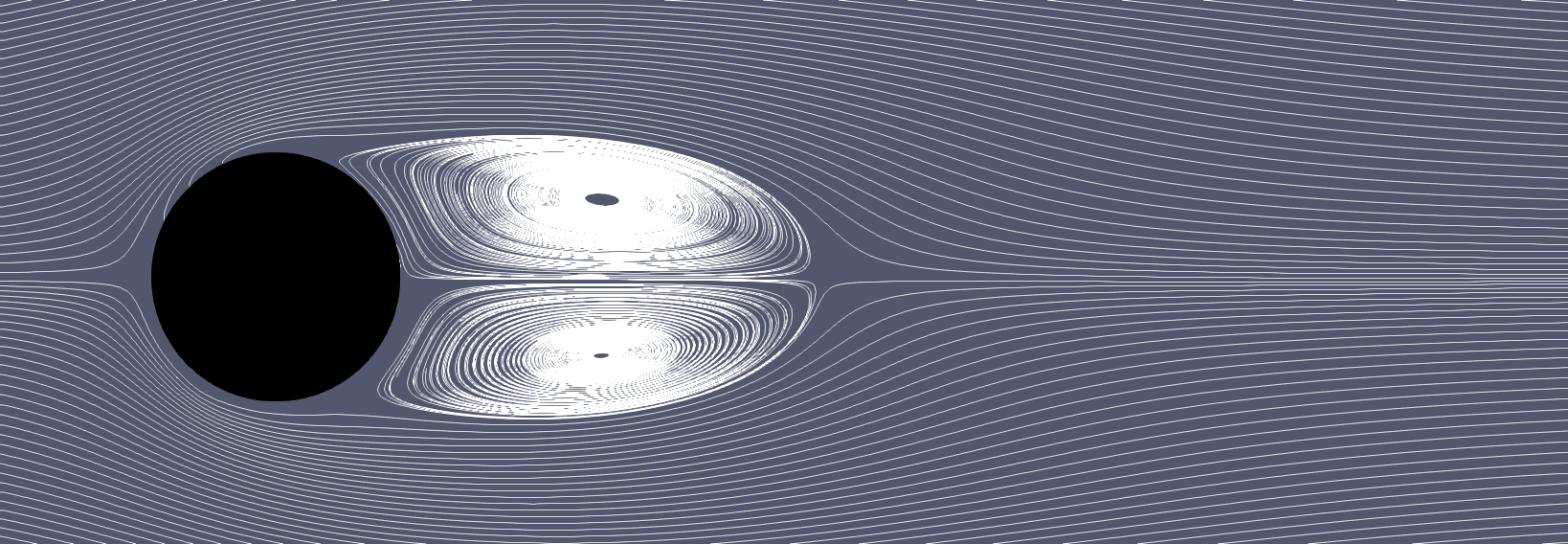}
\caption{Streamline  $(Wi=1.0)$ }
\label{fig:subim8_2}
\end{subfigure}
\begin{subfigure}{0.5\textwidth}
\centering
\includegraphics[width=1.0\linewidth, height=3.0cm]{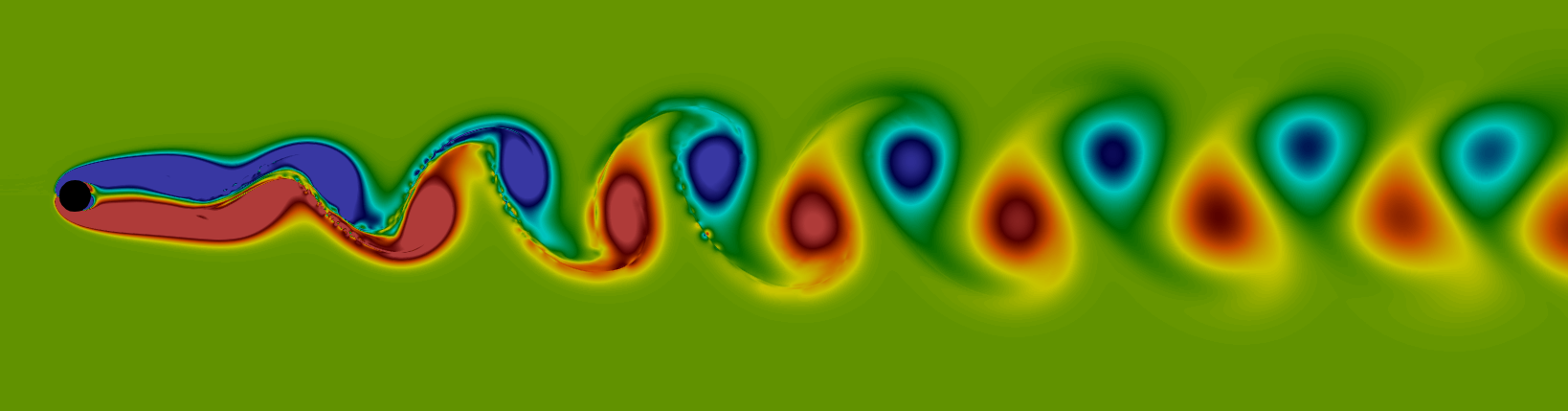}
\caption{Vorticity $\zeta_{z}$ $(Wi=4.0)$ }
\label{fig:subim8_3}
\end{subfigure}
\begin{subfigure}{0.5\textwidth}
\centering
\includegraphics[width=1.0\linewidth,height=3.0cm]{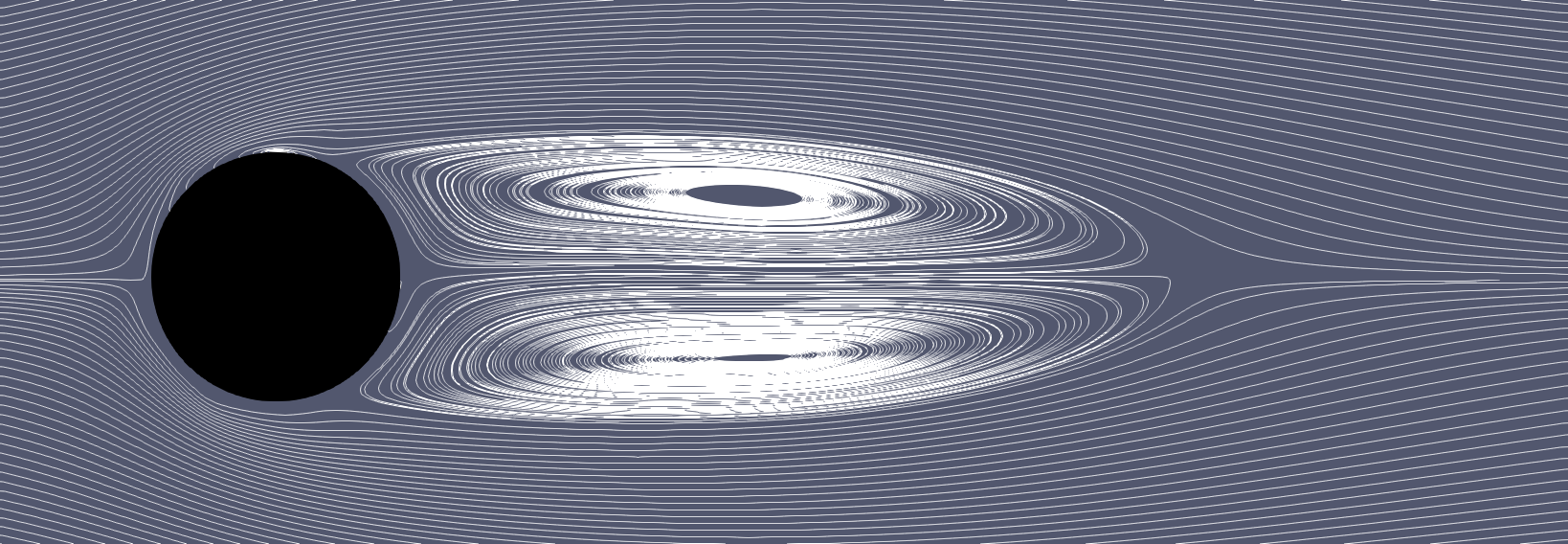}
\caption{Streamline  $(Wi=4.0)$ }
\label{fig:subim8_4}
\end{subfigure}
\caption{The vorticity $\zeta_{z}$ and the streamlines (right) for the viscoelastic fluid flow past a cylinder, for $Wi=1.0$ and 4.0 (top and bottom) and $\beta_{s}=0.9,n=1.0, Bn=0.0$ at $Re=100$. The color scale of the two contour plots is the same.}
\label{fig:image8}
\end{figure}

Elasticity is found to change the vorticity pattern. 
As shown in \Figref{fig:subim8_1} 
the modulations are weak for $Wi=1$ and the vortex shape is similar to the pattern of the Newtonian fluid flow at $Re=100$. 
The double-edged curved dagger -- a "Haladie" shape --vortex appears close to the cylinder for $\beta_{s}=0.9$ and $Wi=1$ and the overall vortex configuration remains close to Newtonian fluid; with the vortices weakening in the streamwise direction. 
Increasing the  elasticity level, by increasing the Weissenberg number to $Wi=4$ or by increasing the polymer concentration by decreasing  the viscosity ratio to $\beta_{s}=0.5$,
the two vortices close to the cylinder extend in the streamwise direction while the bubble length increases.
A comparison of figure~\ref{fig:subim8_1} and \ref{fig:subim8_3} shows that the number of vortices decreases considerably from 17 to 14, as well as their strength, which shows that vortex formation is affected at higher elasticity. However, at the elasticity levels considered here ($\beta_{s}=0.5, Wi=1$ and $\beta_{s}=0.9, Wi=4$), the elastic stresses are not strong enough to prevent the formation of the unsteady \vonK vortex street. 

Figures \ref{fig:subim8_2} and \ref{fig:subim8_4} display the time-averaged streamlines in the presence of elasticity. 
Compared with the Newtonian case in \Figref{fig:subim2_2}, we note that the length of the recirculation bubble increases considerably with elasticity. In addition, in both cases
the center of the recirculation region moves downstream. 
These observations are consistent with previous numerical and experimental studies \cite{Oliveira2001,Peng2021}. The length of the recirculation bubble $L_{RB}$ increases from 1.67 for $\beta_{s}=0.9, Wi=1$ to 3.09 for $\beta_{s}=0.9, Wi=4$. It was also reported in previous numerical and experimental studies, see \cite{Oliveira2001,Oliveira2005,Coelho2003I,Coelho2003II,Richter2010,Peng2021}, that the presence of polymers in the fluid postpones the onset of the \vonK vortex street and decreases the frequency of the vortex shedding,  $St$, considerably. It must be noted that, qualitatively, the vorticity structure remains close to that of a Newtonian fluid flow past a cylinder for the range of non-dimensional numbers of the present study. As mentioned, the maximum polymer stresses occur close to the cylinder surface. However, comparing Figs. \ref{fig:image6} and \ref{fig:image8} also reveals that local maxima of the polymer stresses can be found at the outer edge of strong vorticity regions far away downstream of the cylinder.

It it worth observing that the longer recirculation region behind the cylinder is probably a nonlinear consequence of the suppressed instability and vortex formation, rather than its cause. In global linear instability studies of Newtonian \cite{Juniper2011,Tammisola2011} and Carreau fluids \cite{Iman2012}, a longer recirculation bubble of the based flow increases the growth rates of the instability. In the nonlinear regime however, a high growth rate and a high oscillation amplitude in its turn affects the mean flow and decreases the length of the mean flow recirculation bubble \cite{Tammisola2011}. Similarly, a lower instability amplitude would increase the length of the mean flow recirculation bubble. 

\subsection{The combined effects of viscoelasticity and plasticity}\label{section:Saramito_HB_Model}

In this section, we present the main novelty of the work and 
 focus on the effect of yield stress, shear-thinning, and shear-thickening on the flow structures. 
 The Saramito model \cite{Saramito2009} is used to describe an EVP fluid with shear rate-dependent viscosity in our numerical simulations. \Tabref{tab:3} displays the effect of the yield stress on the flow features for constant $Re=100$ and $Wi=1$. The power-law index is $n=1$, therefore the model is equivalent to the original Saramito model in \cite{Saramito2007}. Comparing  the data in Tables \ref{tab:2} and \ref{tab:3}, we can see that increasing $Bn$ slightly increases $C_D$ for dilute solution, however, for $\beta_{s}=0.5$, the mean of drag coefficient increment is about $10\%$. $C_{L_{rms}}$, $St$, and $r$ decrease with increasing $Bn$, while, $\Theta_{s}$  and $L_{RB}$ increase. As seen, increasing $Bn$ (\ie plasticity) also suppresses the instability; it effectively enhances the effect of elasticity on flow features, as summarized in \Tabref{tab:3}.

\begin{table} 
   \begin{center}
 \def~{\hphantom{0}}
   \begin{tabular}{lccccccc}
   $ Case $               &$C_{D}$ &$C_{L_{rms}}$ & $St$ &$\Theta_{s}$ &$L_{RB}$ &   r   \\[2pt]\hline
   Shear-thinning         &        &              &      &               &                 &       \\[2pt]\hline
   $\beta_{s}=0.9,n=0.2$  &        &              &      &               &                 &       \\[2pt]     
   $Bn=0.0$               &1.43~  &   0.189~     &0.158~&    ~115.5~    &     1.67~       &35.7~  \\     
   $Bn=1.0$               &1.27~  &   0.133~     &0.160~&    ~117.5~    &     1.75~       &39.7~  \\     
   $Bn=2.0$               &1.24~  &   0.073~     &0.156~&    ~118.1~    &     2.11~       &57.5~  \\\hdashline
   $\beta_{s}=0.5,n=0.2$  &        &              &      &               &                 &       \\[2pt]     
   $Bn=0.0$               &1.52~  &   0.073~     &0.14~ &    ~108.7~    &     2.56~       &~5.2~  \\     
   $Bn=2.0$               &1.11~  &   0.039~     &0.163~ &    ~111.8~    &     2.98~       &~8.6~  \\\hdashline 
    $\beta_{s}=0.9,n=0.6$  &        &              &      &               &                 &       \\[2pt]     
   $Bn=0.0$               &1.43~  &   0.189~     &0.157~&    ~116.3~    &     1.66~       &35.8~  \\     
   $Bn=1.0$               &1.28~  &   0.103~     &0.161~&    ~119.1~    &     1.89~       &49.2~  \\     
   $Bn=2.0$               &1.25~  &   0.052~     & ~-~  &    ~119.7~    &     2.40~       &29.5~  \\\hline    
   $n=1$         &        &              &      &               &                 &       \\\hline
   $\beta_{s}=0.9$        &        &              &      &             &                   &       \\[2pt]     
   $Bn=1.0$               &1.440~  &   0.034~     &0.139~&    ~114.9~  &     2.95~         &11.7~  \\     
   $Bn=2.0$               &1.438~  &   0.022~     &  ~-~ &    ~116.1~  &     3.75~         &~3.5~  \\\hdashline
   $\beta_{s}=0.5$        &        &              &      &             &                   &       \\[2pt]     
   $Bn=2.0$               &1.665~  &   0.007~     &  ~-~ &    ~114.1~  &     4.67~         &~0.9~  \\\hline  
   Shear-thickening       &        &              &      &             &                   &       \\\hline
   $\beta_{s}=0.9,n=1.4$  &        &              &      &             &                   &       \\[2pt]     
   $Bn=0.0$               &1.432~  &   0.189~     &0.157~&    ~115.5~  &     1.66~         &35.9~  \\     
   $Bn=1.0$               &1.887~  &   0.114~     &  ~-~ &   ~121.9~   &     3.42~         &~1.0~  \\     
   $Bn=2.0$               &2.00~  &   0.08~     &  ~-~ &   ~127.3~   &     3.61~         &~0.7~  \\\hdashline     
   $\beta_{s}=0.9,n=2.0$  &        &              &      &             &                   &       \\[2pt]     
   $Bn=0.0$               &1.432~  &   0.189~     &0.157~&    ~115.4~  &     1.66~         &35.9~  \\     
   $Bn=1.0$               &2.068~  &   0.196~     &  ~-~ &   ~123.3~   &     1.84~         &~0.9~  \\     
   $Bn=2.0$               &2.153~  &   0.23~     &  ~-~ &   ~124.7~   &     1.15~         &~1.0~  \\\hdashline
   $\beta_{s}=0.5,n=2.0$  &        &              &      &             &                   &       \\[2pt]     
   $Bn=0.0$               &1.520~  &   0.072~     &0.142~&   ~110.3~   &     2.54~         &~5.0~  \\     
   $Bn=2.0$               &2.05~  &   0.172~     &  ~-~ &   ~120.6~   &     2.01~         &~0.9~  \\\hline    
   \end{tabular}
   \caption{Flow characteristics (the drag coefficient $C_D$, the rms of lift coefficient $C_{L_{rms}}$, Strouhal number $St$, separation angle $\Theta_{s}$, the recirculation bubble length $L_{RB}$, and the ratio of fluctuations of drag and lift coefficients \Pab{$r=(\frac{C_{L_{rms}}}{C_{D_{rms}}})$} for the elastoviscoplastic flow past a circular cylinder at Reynolds number $Re=100$ and Weissenberg number $Wi=1.0$,  and the values of the power law index $n$, Bingham number $Bn$ and viscosity ratio $\beta_{s}$ indicated. }
   \label{tab:3}
   \end{center}
 \end{table}

\begin{figure} 
\begin{subfigure}{0.5\textwidth}
\centering
\includegraphics[width=0.85\linewidth, height=5.5cm]{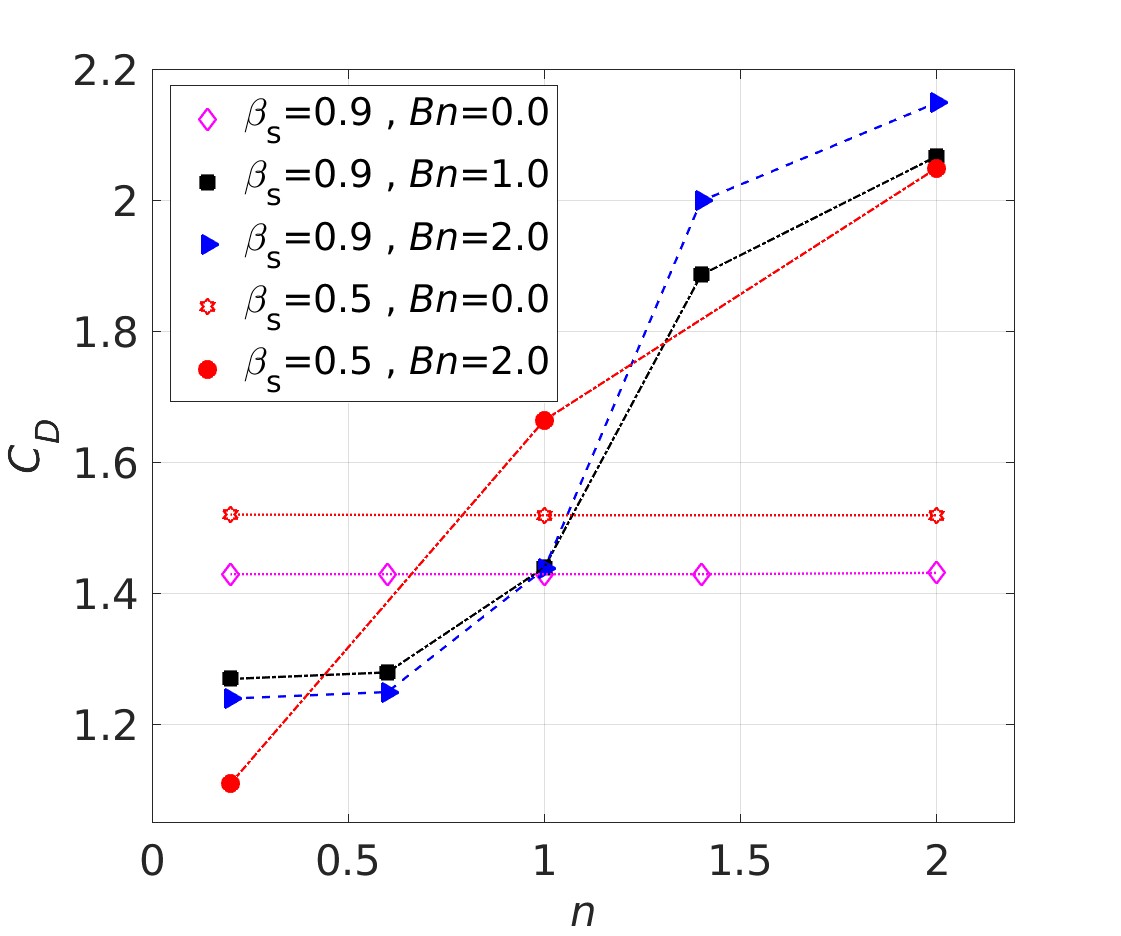}
\caption{\label{EVP_Cd} $C_D$ }
\label{fig:subim9_1}
\end{subfigure}
\begin{subfigure}{0.5\textwidth}
\centering
\includegraphics[width=0.95\linewidth, height=5.5cm]{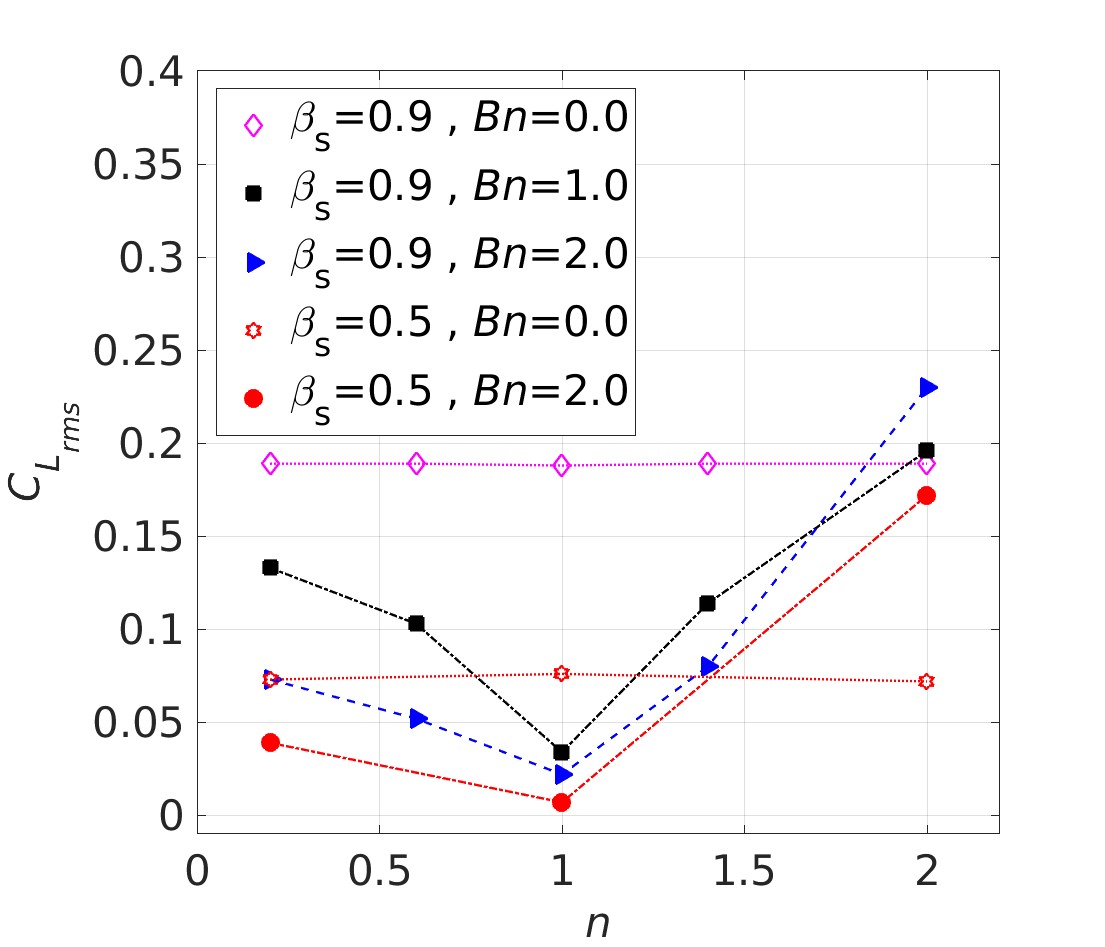}
\caption{\label{EVP_CL} $C_{L_{rms}}$ }
\label{fig:subim9_2}
\end{subfigure}
\caption{(a) Mean drag coefficient and (b) root mean square of the lift coefficient of a viscoelastic and EVP fluid with $n=[0.2,0.6,1.0,1.4,2.0]$, $Bn=[0.0,1.0,2.0]$, and $\beta_{s}=0.9,0.5$, at  $Re=100$ (the viscoelastic cases are shown by open symbols and EVP flows by solid symbols.}
\label{fig:image9}
\end{figure}

\Tabref{tab:3} also reports the effect of shear-thinning and shear-thickening on the elastoviscoplastic flow features for constant $Re =100$, $Wi=1$ and different values of $Bn$. 
Also for the shear-thinning fluid, $n=0.6$ and $n=0.2$,  $C_{L_{rms}}$ decreases if $Bn$ increases, while $L_{RB}$ and $\Theta_{s}$ increase. However, these effects are smaller than in the shear-independent case, and $C_{L_{rms}}$ remains 10 times larger ($0.007$ when $n=1$, and $0.073$ when $n=0.2$). This indicates that shear-thinning weakens the combined stabilizing effects of elastic and plastic forces.
The reduction of $C_{D}$ and $C_{L_{rms}}$ in shear thinning fluid with $Bn>0$ is depicted in detail in \Figref{fig:image9}. 
The effect of shear-thinning on the shedding frequency $St$ and the fluctuation ratio $r$ is complicated and non-monotonic in the range considered in the present study. 
For shear-thickening fluid, increasing $Bn$ increases $C_{D}$ as seen in \Figref{fig:image9}, and noticeably decreases $St$ and the $r$ ratio, a sign of change to an irregular flow pattern.
It is worth mentioning that we did not identify any dominant frequency in the case of a shear-thickening fluid in the presence of yield stress, and that is why the entries for $St$ in \Tabref{tab:3} are empty. 
The effect of yield stress and shear thickening on $L_{RB}$ is complex. For instance, for $n=1.4$, increasing $Bn$ lengthens the recirculation region. However, for an even more shear-thickening fluid ($n=2.0$), increasing $Bn$ shortens the recirculating bubble. The main reason behind this contradiction is the change of flow pattern from periodic to chaotic, shown by the considerable increase of $C_{L_{rms}}$ for ($Bn>0$ and $n=2.0$) in \Figref{fig:image9}, which will be further  discussed later. A more chaotic flow leads to stronger nonlinear effects that affect the mean flow and shorten the recirculation bubble.
Decreasing $\beta_{s}$ from values of a dilute solution ($\beta_{s}=0.9$) to concentrated polymer solutions ($\beta_{s}= 0.5$) for the viscoelastic case (\ie $Bn = 0$ and 
$n \ne 1$) increases $C_{D}$, as shown in \Figref{fig:image9}, and moderately decreases $St$ in both shear thinning and thickening fluid with $n=0.2$ and $n=2$, see  \Tabref{tab:3}. However, when the yield stress becomes significant at $Bn=2$, $C_{D}$ decreases, and $St$ slightly increases with decreasing $\beta_s$. It is noteworthy that by decreasing $\beta_{s}$, the length of the recirculation bubble increases.

\begin{figure} 
\begin{subfigure}{0.33\textwidth}
\centering
\includegraphics[width=0.9\linewidth, height=4cm]{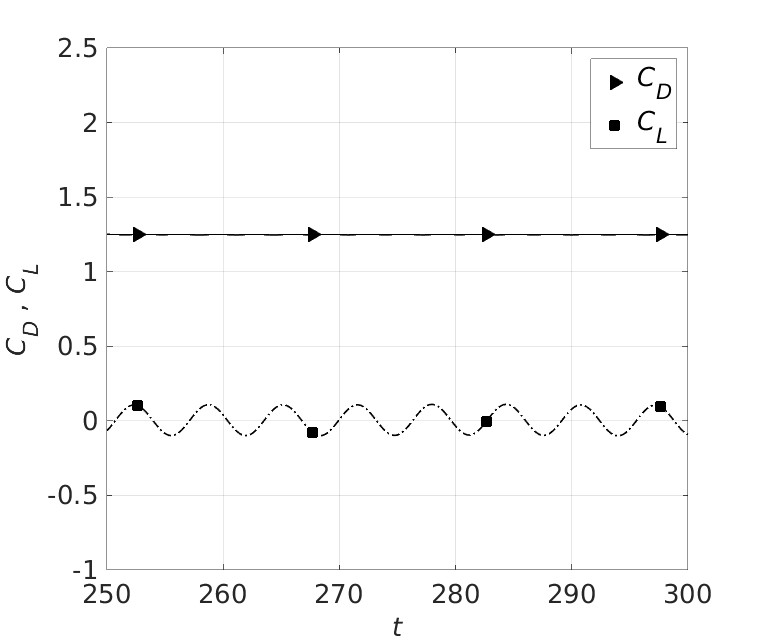}
\caption{Shear thinning $n=0.2$}
\label{fig:subim10_1}
\end{subfigure}
\begin{subfigure}{0.33\textwidth}
\centering
\includegraphics[width=0.9\linewidth, height=4cm]{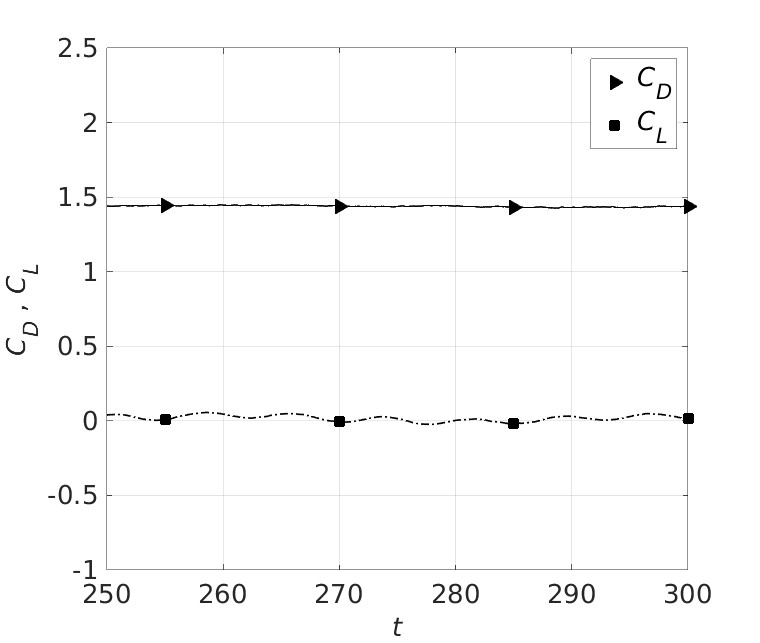}
\caption{shear-independent $n=1.0$}
\label{fig:subim10_2}
\end{subfigure}
\begin{subfigure}{0.33\textwidth}
\centering
\includegraphics[width=0.9\linewidth, height=4cm]{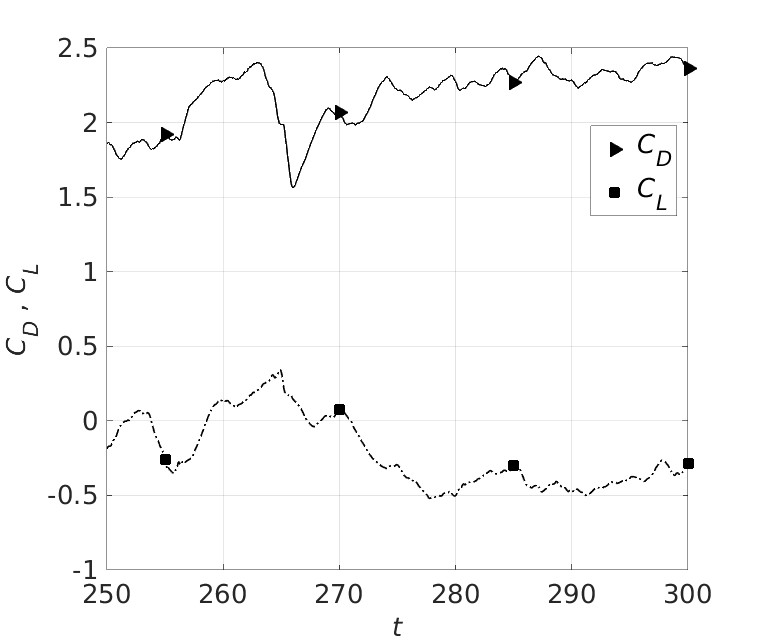}
\caption{Shear thickening $n=0.2$}
\label{fig:subim10_3}
\end{subfigure}
\caption{Drag and lift coefficient time evolution  for an EVP fluid with $Wi=1.0$, $\beta_{s}=0.9$, $Bn=2.0$ and at $Re=100$, for (a) a shear thinning ($n=0.2$), (b) shear-independent $n=1.0$ and (c) shear-thickening ($n=2.0$).}
\label{fig:image10}
\end{figure}

As shown in \Tabref{tab:3}, for $n=1$, in the presence of yield stress, the amplitude of $C_L$ decreases with increasing elasticity (decreasing $\beta_{s}$) and $Bn$. 
The time evolution of the drag and lift coefficients, $C_D$ and $C_L$, is depicted in \Figref{fig:image10}
 for different values of the index $n$, $n=[0.2,1.0,2.0]$, at constant $\beta_{s}=0.9$, $Wi=1$ and $Bn=2$. for the specific range of non-dimensional time between $t=250-300$. For shear-thinning fluid with $n=0.2$ and dilute polymer solution $\beta_{s}=0.9$, the sinusoidal behaviour can be clearly seen in \Figref{fig:subim10_1}.
 For a shear-thickening fluid with $n=2$, the flow pattern is no longer time-periodic, but becomes irregular and reminiscent of elastic instability, due to 
  the combination of shear-thickening, yield stress, and elasticity. 
  In the following, we show that these irregular fluctuations totally change the flow structure downstream of the cylinder. We want to emphasize that although lift coefficients seem negative for the shear thickening flow in the time interval shown, their time average over the entire simulation is close to zero. For concentrated solutions with $\beta_{s}=0.5$,  and shear-thinning $n=0.2$, or $n=1$, the amplitude of $C_L$ is considerably reduced. For the shear-thickening case, instead, strong and irregular fluctuations in $C_D$ and $C_L$ are observed.

\begin{figure} 
\begin{subfigure}{0.33\textwidth}
\centering
\includegraphics[width=0.9\linewidth, height=4cm]{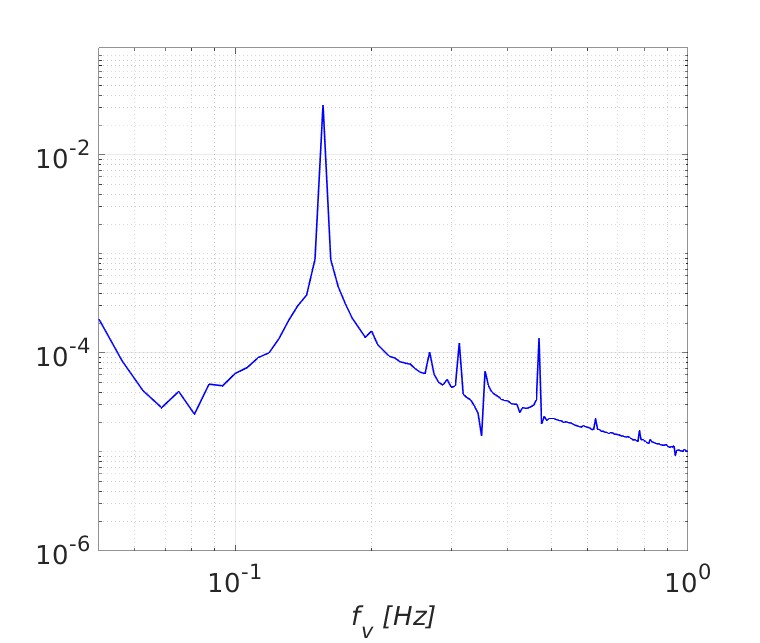}
\caption{\label{FFT_Re100_Wi1.0_n0.2_Bs0.9_Bn2.0} Shear thinning $n=0.2$ }
\label{fig:subim11_1}
\end{subfigure}
\begin{subfigure}{0.33\textwidth}
\centering
\includegraphics[width=0.9\linewidth, height=4cm]{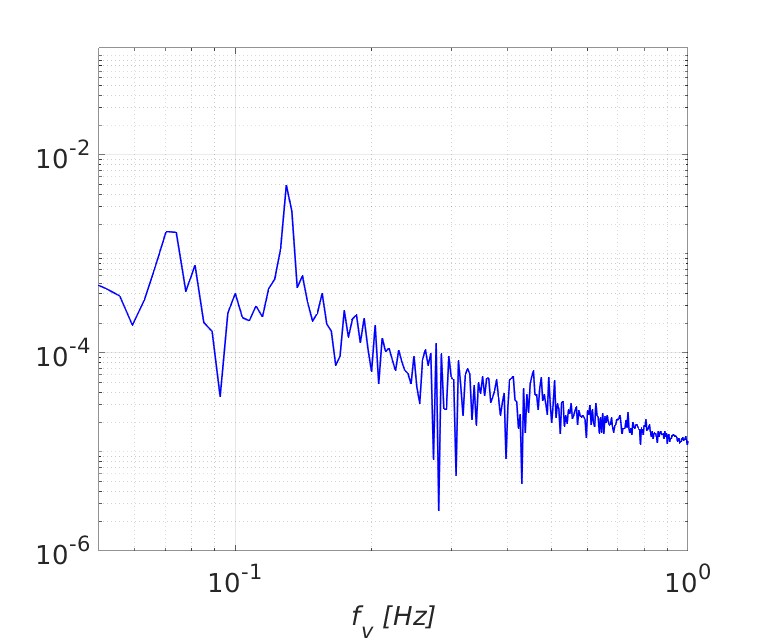}
\caption{\label{FFT_Re100_Wi1.0_n1.0_Bs0.9_Bn2.0} shear-independent $n=1.0$ }
\label{fig:subim11_2}
\end{subfigure}
\begin{subfigure}{0.33\textwidth}
\centering
\includegraphics[width=0.9\linewidth, height=4cm]{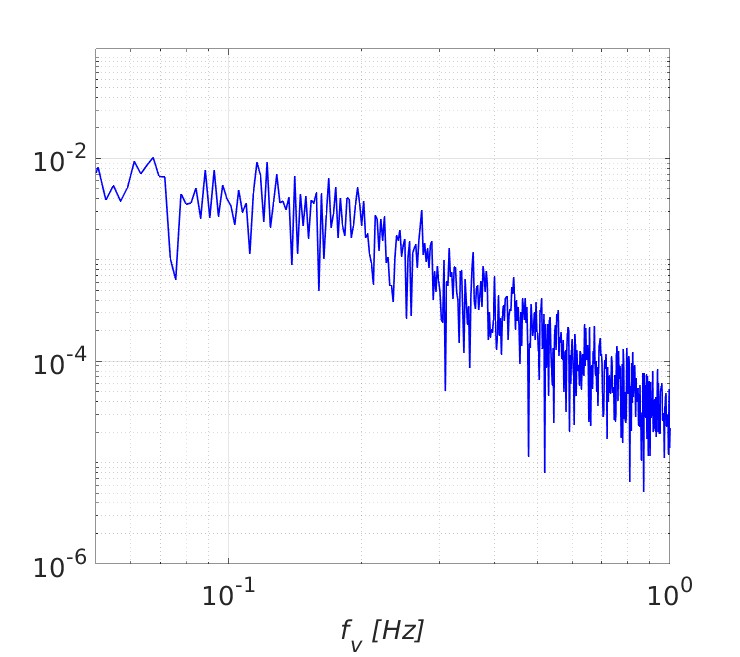}
\caption{\label{FFT_Re100_Wi1.0_n2.0_Bs0.9_Bn2.0} Shear thickening $n=2.0$ }
\label{fig:subim11_3}
\end{subfigure}
\caption{ The frequency $f_{v}$ for 
for an EVP fluid with $Wi=1.0$ ,$\beta_{s}=0.9$, $Bn=2.0$ and at $Re=100$, for (a) a shear thinning ($n=0.2$), (b) $n=1.0$ and (c) shear-thickening ($n=2.0$).}
\label{fig:image11}
\end{figure}

\Figref{fig:image11} show the frequency for a dilute EVP fluid with $\beta_{s}=0.9, Wi=1.0$, at $Re=100$, and different shear-dependent viscosity, i.e.\ shear thinning ($n=0.2$), shear-independent $n=1$, and shear thickening ($n=2.0$). For shear thinning ($n=0.2$) and shear-independent ($n=1$) flows, we see one dominant frequency. The peak amplitude is largest in the $n=0.2$, while at $n=1$ the peak is lower and lower frequencies start to become almost comparable. However, we could not obtain any dominant frequency for the shear-thickening ($n=2.0$) fluid, in which case the flow changes from periodic to chaotic. It is interesting to observe that the $n=1$ spectrum is in between the two extreme cases, but we still observe a very regular vortex pattern in figure \ref{fig:subim11_2}.  

\begin{figure} 
\begin{subfigure}{0.6\textwidth}
\centering
\includegraphics[width=1.0\linewidth, height=2.85cm]{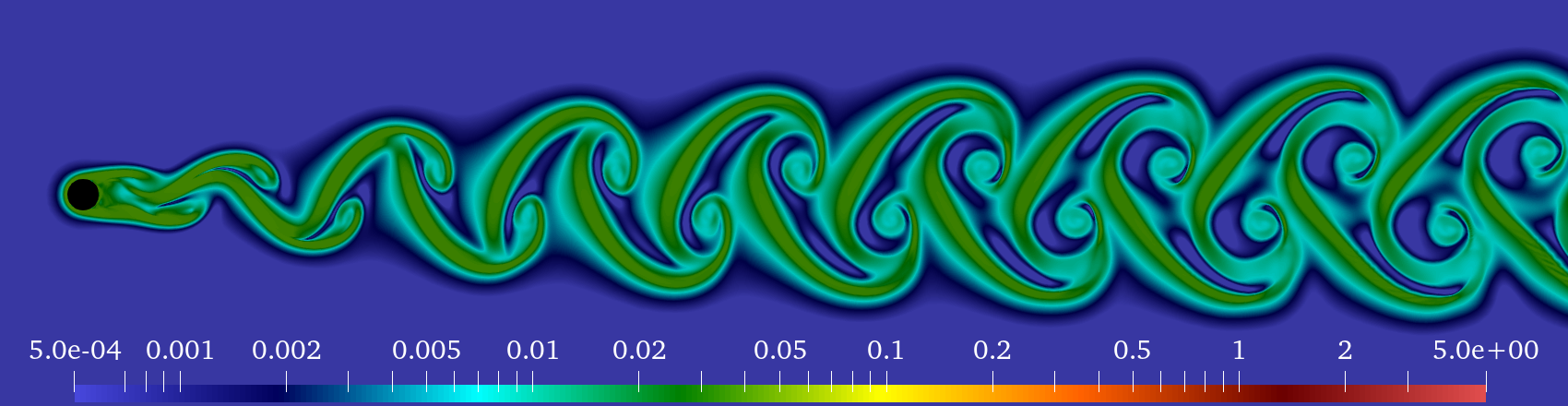}
\caption{Shear thinning $n=0.2$ }
\label{fig:subim12_1}
\end{subfigure}
\begin{subfigure}{0.4\textwidth}
\centering
\includegraphics[width=1.0\linewidth, height=2.85cm]{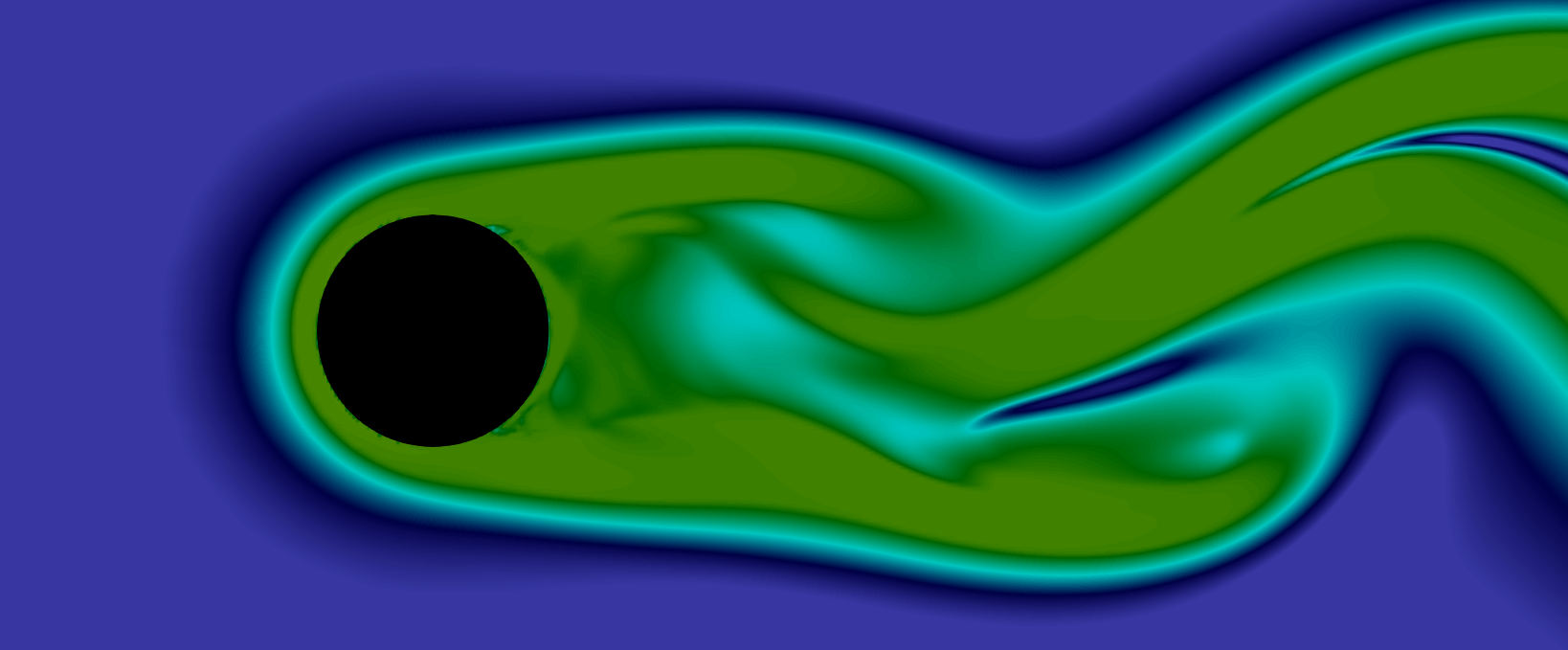}
\caption{The zoom of \Figref{fig:subim12_1}}
\label{fig:subim12_2}
\end{subfigure}
\begin{subfigure}{0.6\textwidth}
\centering
\includegraphics[width=1.0\linewidth, height=2.85cm]{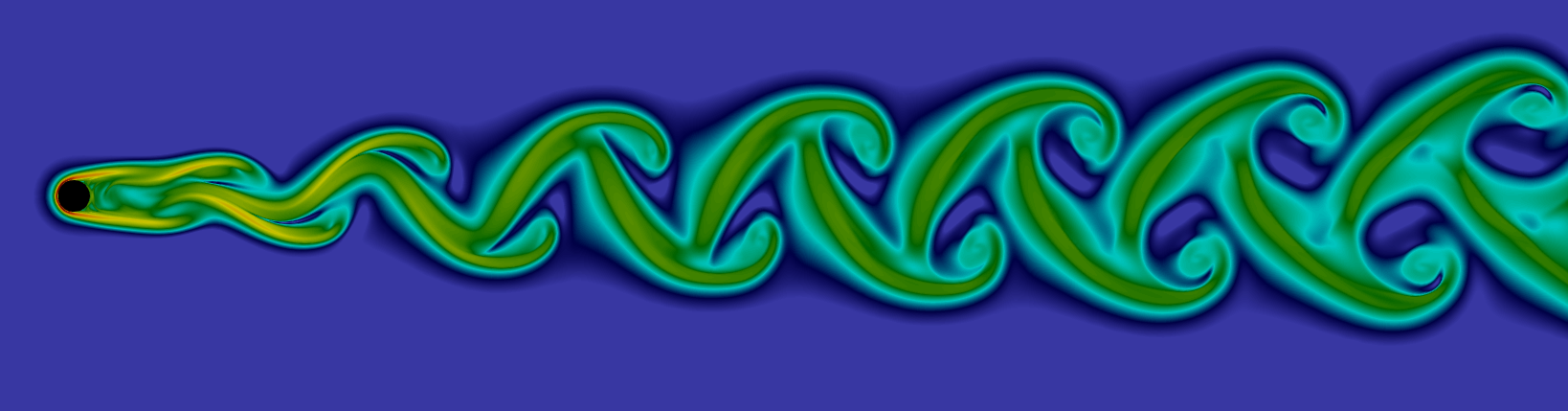}
\caption{shear-independent $n=1.0$}
\label{fig:subim12_3}
\end{subfigure}
\begin{subfigure}{0.4\textwidth}
\centering
\includegraphics[width=1.0\linewidth, height=2.85cm]{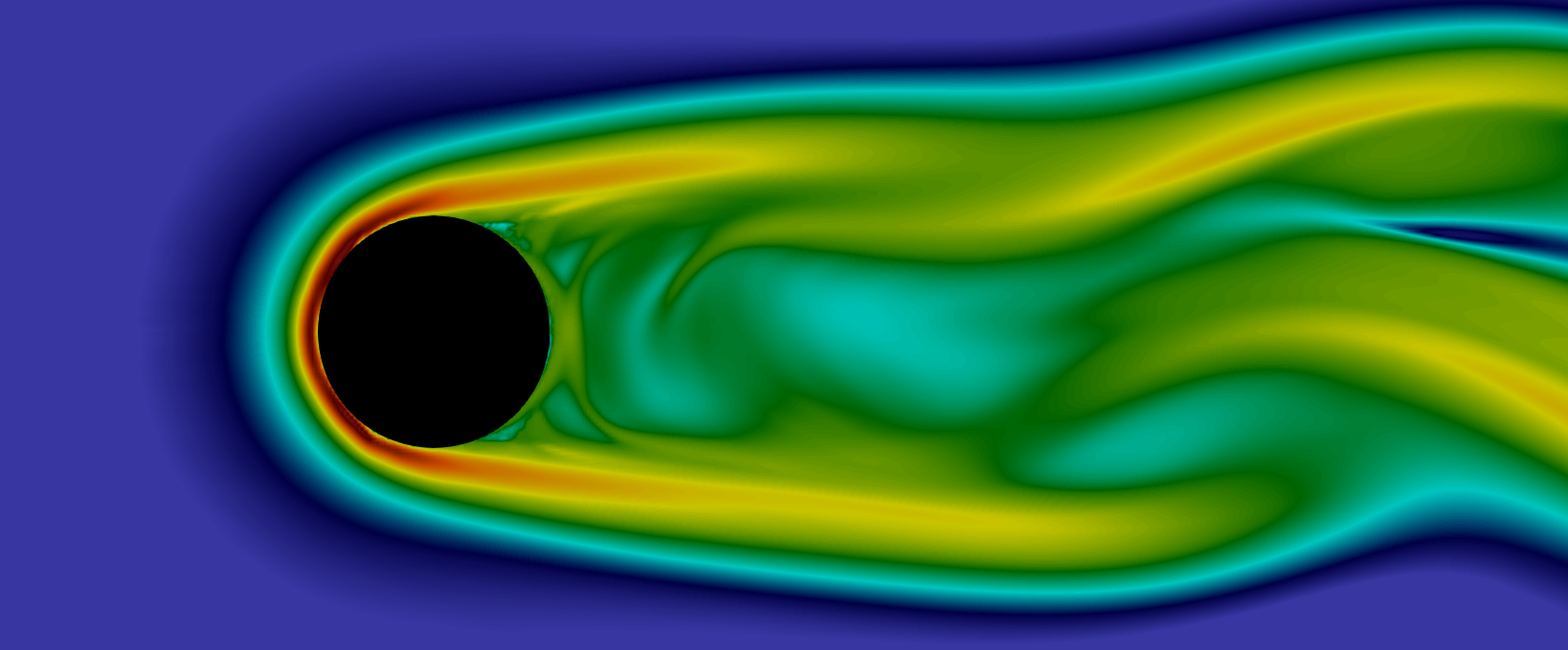}
\caption{The zoom of \Figref{fig:subim12_3}}
\label{fig:subim12_4}
\end{subfigure}
\begin{subfigure}{0.6\textwidth}
\centering
\includegraphics[width=1.0\linewidth, height=2.85cm]{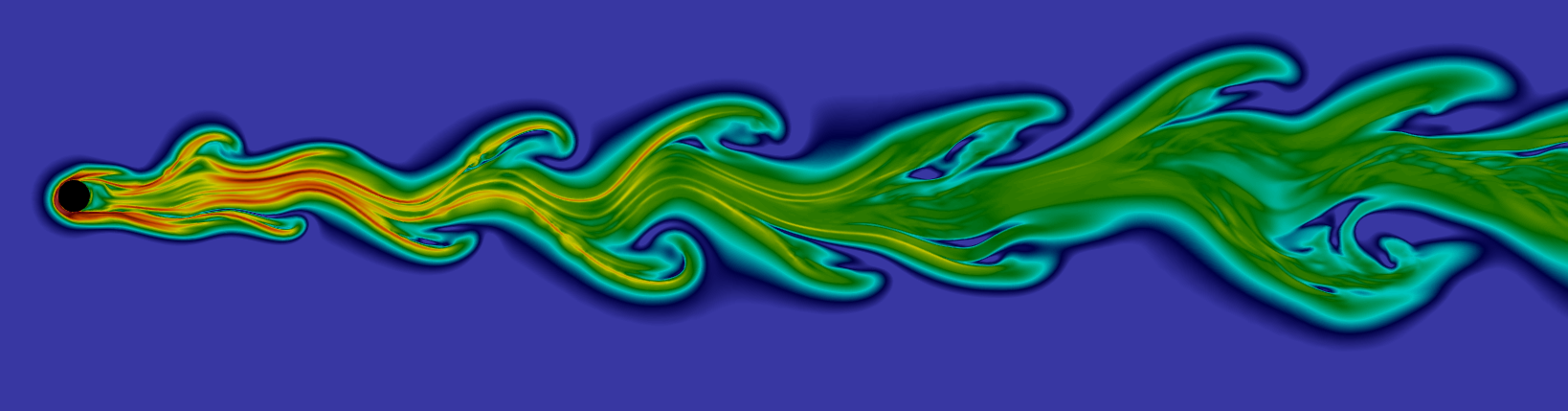}
\caption{Shear thickening$n=2.0$ }
\label{fig:subim12_5}
\end{subfigure}
\begin{subfigure}{0.4\textwidth}
\centering
\includegraphics[width=1.0\linewidth, height=2.85cm]{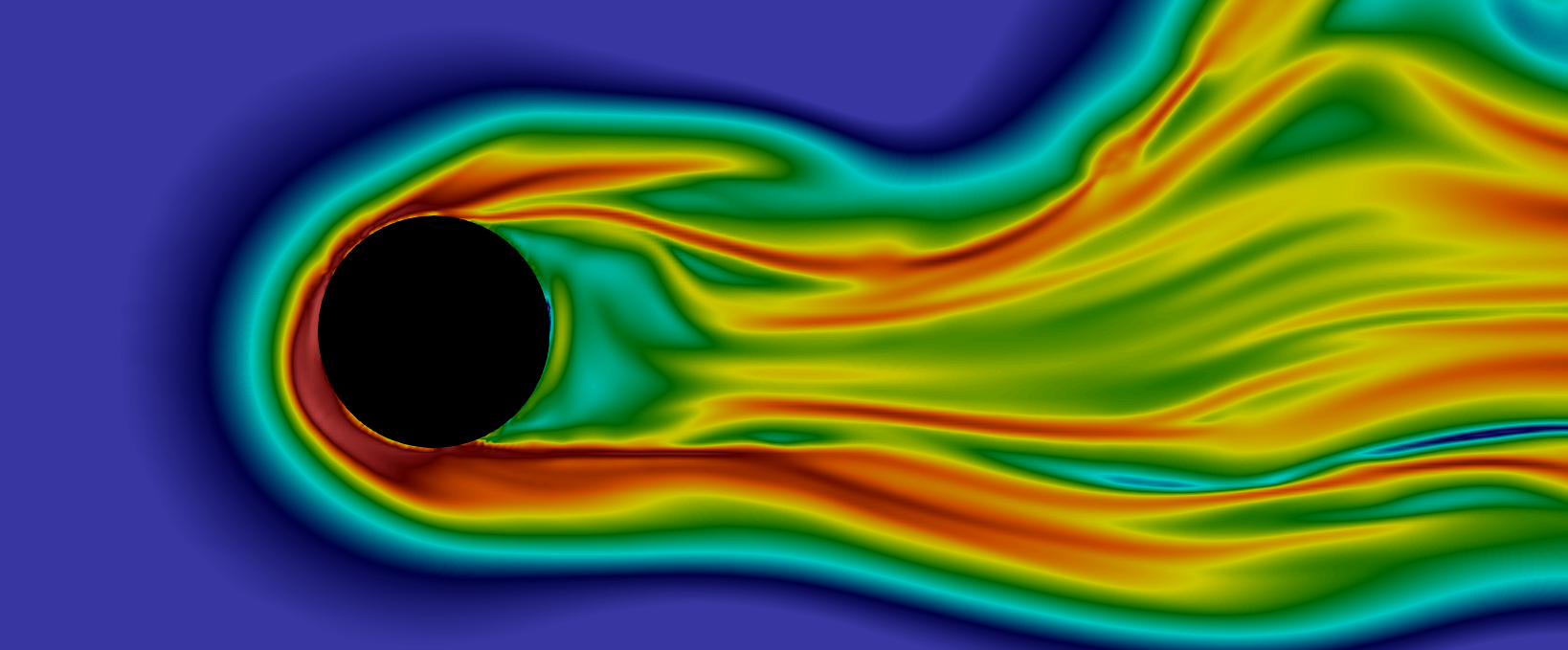}
\caption{The zoom of \Figref{fig:subim12_5}}
\label{fig:subim12_6}
\end{subfigure}
\caption{The trace of polymer stress contour $\trace{\btau^p}=\tau_{xx}^{p}+\tau_{yy}^{p}+\tau_{zz}^{p}$ for variation of elastoviscoplastic fluid with $\beta_{s}=0.9$, $Wi=1.0$, $n=[0.2,1.0,2.0]$, and $Bn=2.0$ at $Re=100$. The color scale is the same in all plots.}
\label{fig:image12}
\end{figure}

Next, we examine the trace of polymer stress tensor $\trace{\btau^p}=\tau_{xx}^{p}+\tau_{yy}^{p}+\tau_{zz}^{p}$ for the EVP fluid with $Bn=2.0$ and various $n=[0.2,1.0,2.0]$, see figure \ref{fig:image12}. Similarly to \Figref{fig:image6}, on the left column of \Figref{fig:image12}, the contours show the trace of the polymer stress tensor distribution over
 the whole domain, while on the right column, we focus on the region around the cylinder. 
For a dilute EVP fluid with $\beta_{s}=0.9$, $Bn=2$, with $n=1$ or shear-thinning $n=0.2$, the polymer stress shows a periodic behaviour. In this condition, the global maximum polymer stress occurs close to the cylinder surface and the local maximum polymer stress exists on the vortex edges far away from the cylinder.
When the yield stress is considerable, \ie $Bn=2$, it can be seen in Figs.\ref{fig:subim12_1}-\ref{fig:subim12_6} that the $\trace{\btau^p}$ decays more slowly 
downstream of the cylinder, indicating that the elastic stretching and stresses persist at large distances downstream from the cylinder when compared to the viscoelastic fluid shown \Figref{fig:image6}. 
This shows that the combination of yield stress and elasticity can affect the flow structures significantly far downstream of the cylinder. The high values of elastic stretching at $Bn=2$ also indicate that the chaotic flow is a result of an elastic instability, even though the elasticity number $EI=0.04 $ is low, since yield stress increases the elastic effects. It is also observed that shear-thinning reduces $\trace{\btau^p}$. Regions with high values of $\trace{\btau^p}$ occur upstream and extend further downstream in the case of a shear-thickening fluid. Also in the case $\beta_{s}=0.5$ and for a shear-thickening fluid ($n=2$), 
the flow becomes chaotic when the yield stress is sufficiently high $Bn=2.0$, although not shown for brevity. 
In this case, although the global maximum of $\trace{\btau^p}$ still occurs close to the cylinder, regions of high polymer extension extends all the way to the centerline. The figure for $\beta=0.5$ is not in the paper. We can either add it as the last subfigure, or remove the sentence. 
Focusing on the zoomed-in section of \Figref{fig:image12} reveals that increasing the power law index $n$ in EVP fluid, 
the polymer stretching is increased in the 
vicinity of the cylinder. For the shear-thickening fluid when the flow is chaotic, the maximum stretching also occurs slightly downstream of the cylinder, \Figref{fig:subim12_6}.

We also studied the strain rate tensor, $\tens{S}(\tens{u})=S_{ij}$, in index notation reads as
\begin{equation}
S_{ij}=\frac{1}{2}\left(\pd{u_i}{x_j}+\pd{u_j}{x_i} \right)
\label{eq:strainRateIndex}
\end{equation}
It is observed that the maximum magnitude of $S_{xx}$ exists on the upstream side of the cylinder, while thick streaks of large $S_{xy}$ contours exist upstream of the cylinder and extend downstream from the cylinder surface. In the presence of yield stress ($Bn=2$), at a large enough distance downstream of the cylinder, the local maximum of all cases exist at the outer layer of the vortex. Generally speaking, the comparison of the strain rate tensor components reveals that the magnitude of $S_{xy}$ is higher than $S_{xx}$ and $S_{yy}$. These contours are also not shown for brevity. 

\begin{figure} 
\begin{subfigure}{0.6\textwidth}
\centering
\includegraphics[width=1.0\linewidth, height=2.85cm]{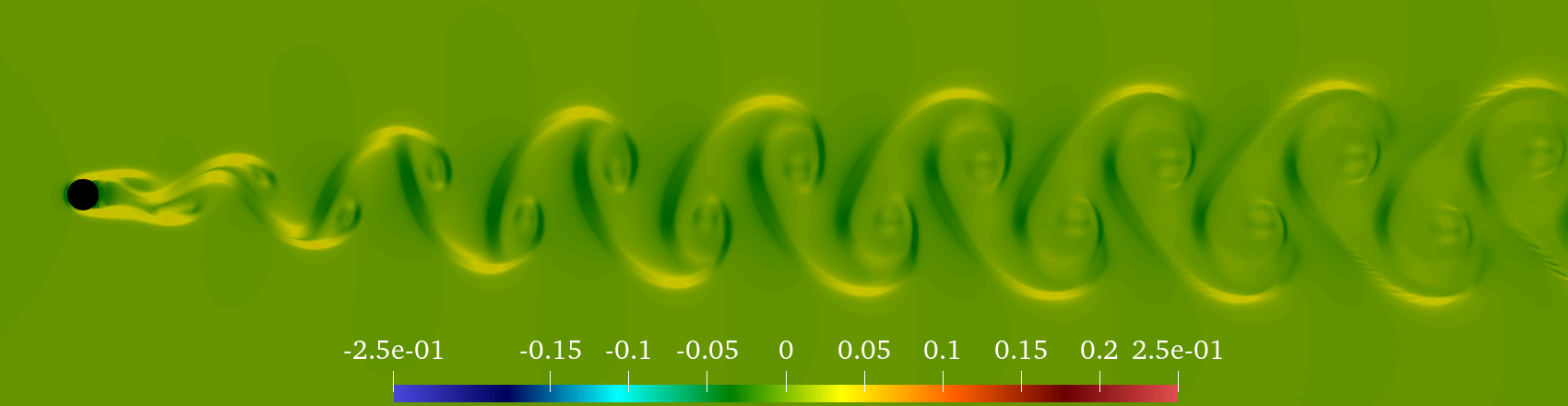}
\caption{Shear thinning $n=0.2$}
\label{fig:subim13_1}
\end{subfigure}
\begin{subfigure}{0.4\textwidth}
\centering
\includegraphics[width=1.0\linewidth, height=2.85cm]{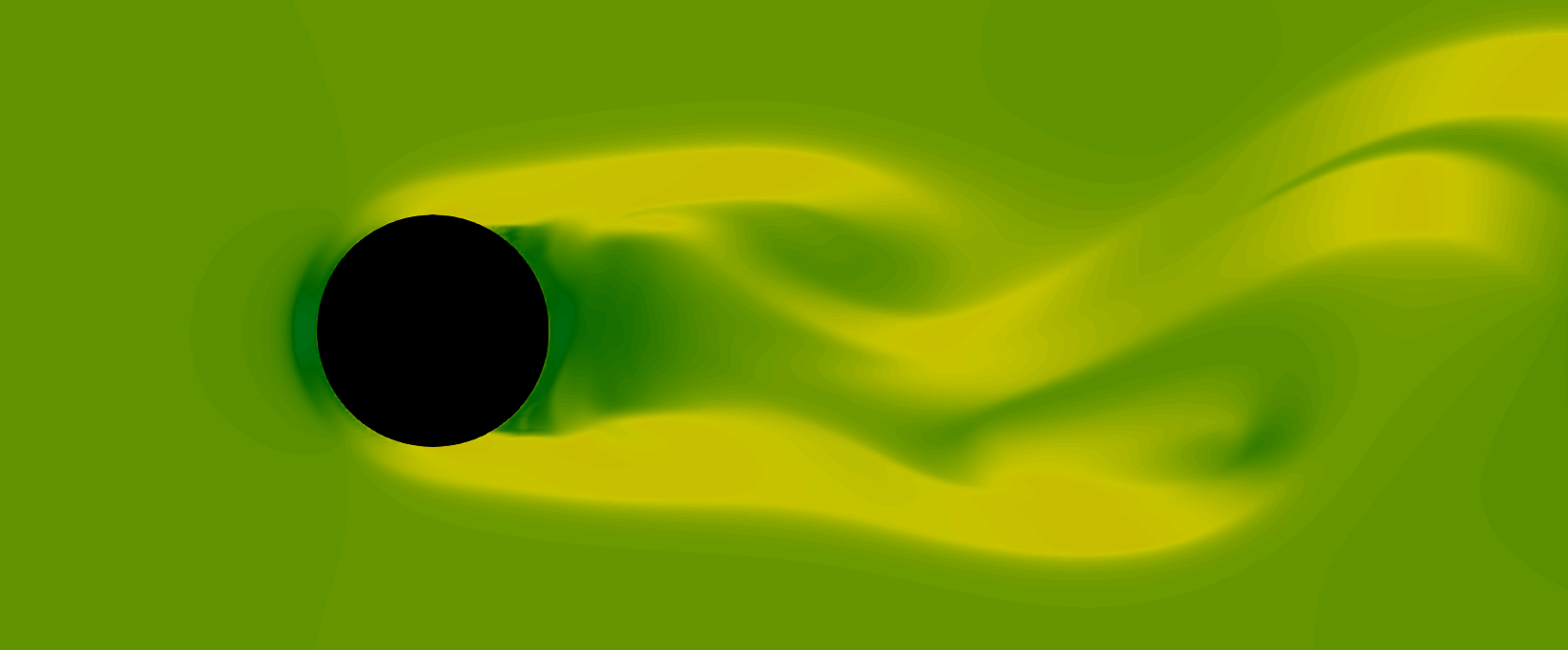}
\caption{Close-up of \Figref{fig:subim13_1}}
\label{fig:subim13_2}
\end{subfigure}
\begin{subfigure}{0.6\textwidth}
\centering
\includegraphics[width=1.0\linewidth, height=2.85cm]{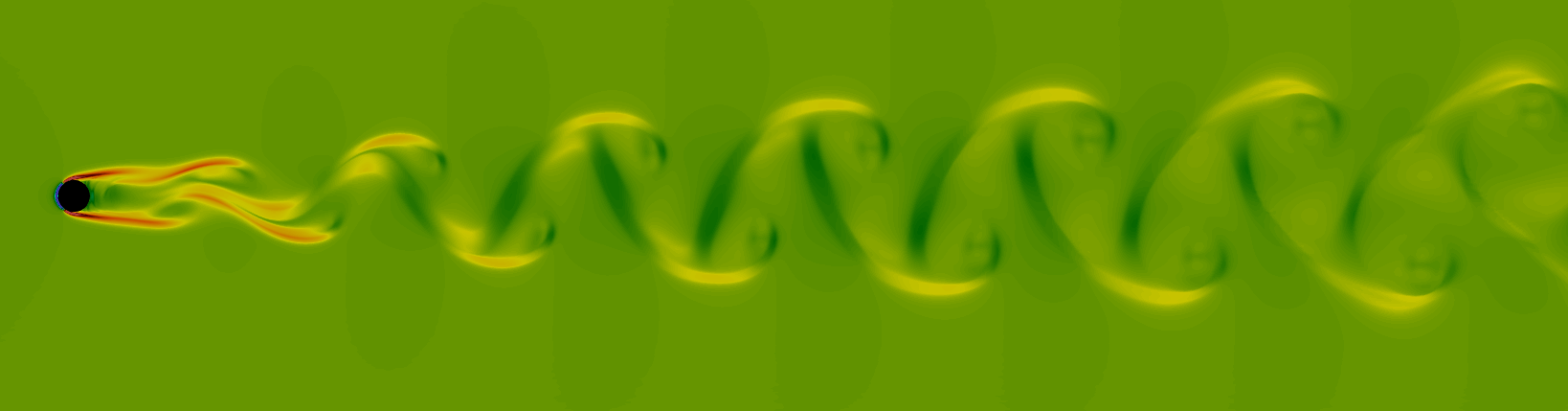}
\caption{shear-independent $n=1.0$ }
\label{fig:subim13_3}
\end{subfigure}
\begin{subfigure}{0.4\textwidth}
\centering
\includegraphics[width=1.0\linewidth, height=2.85cm]{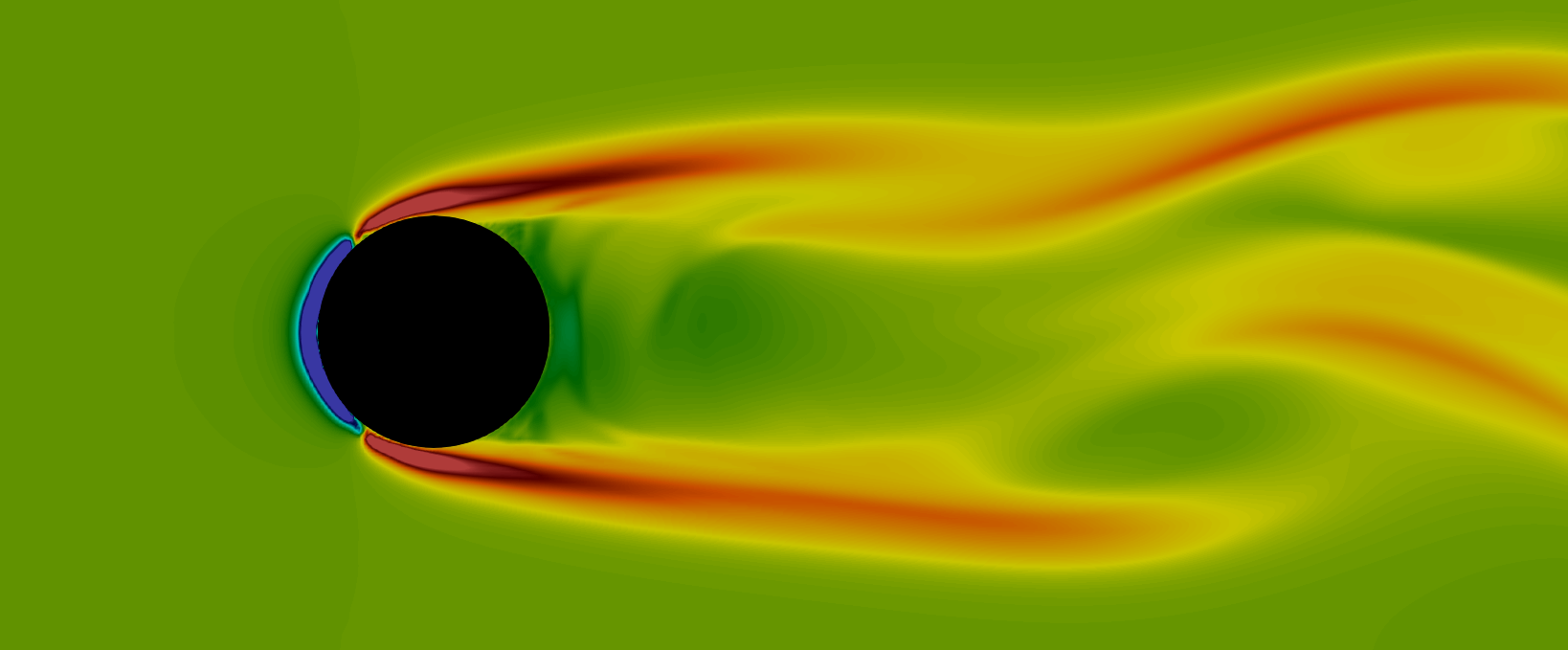}
\caption{Close-up of \Figref{fig:subim13_3}}
\label{fig:subim13_4}
\end{subfigure}
\begin{subfigure}{0.6\textwidth}
\centering
\includegraphics[width=1.0\linewidth, height=2.85cm]{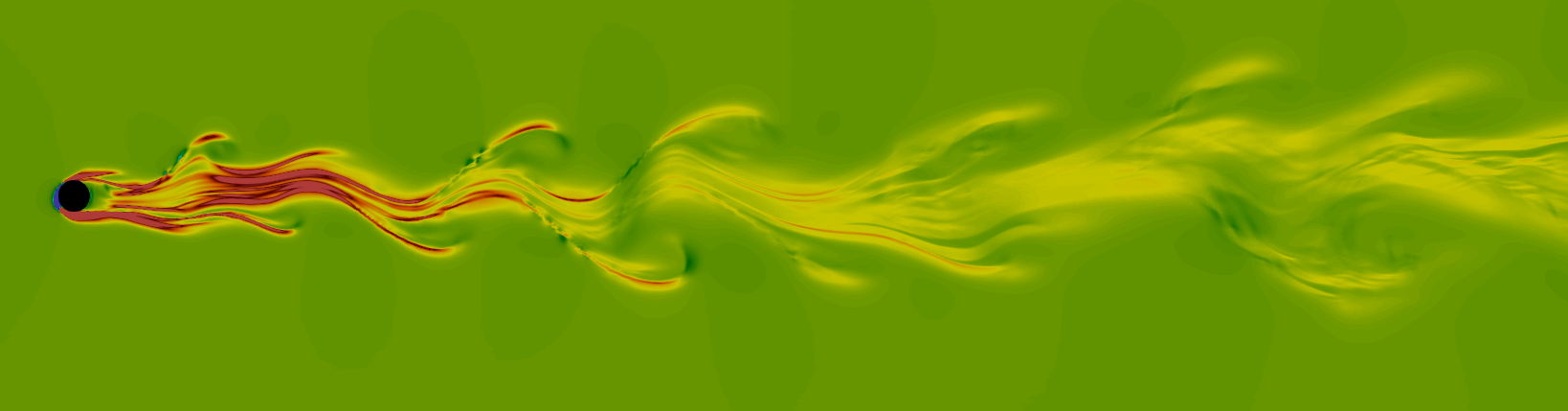}
\caption{Shear thickening $n=2.0$}
\label{fig:subim13_5}
\end{subfigure}
\begin{subfigure}{0.4\textwidth}
\centering
\includegraphics[width=1.0\linewidth, height=2.85cm]{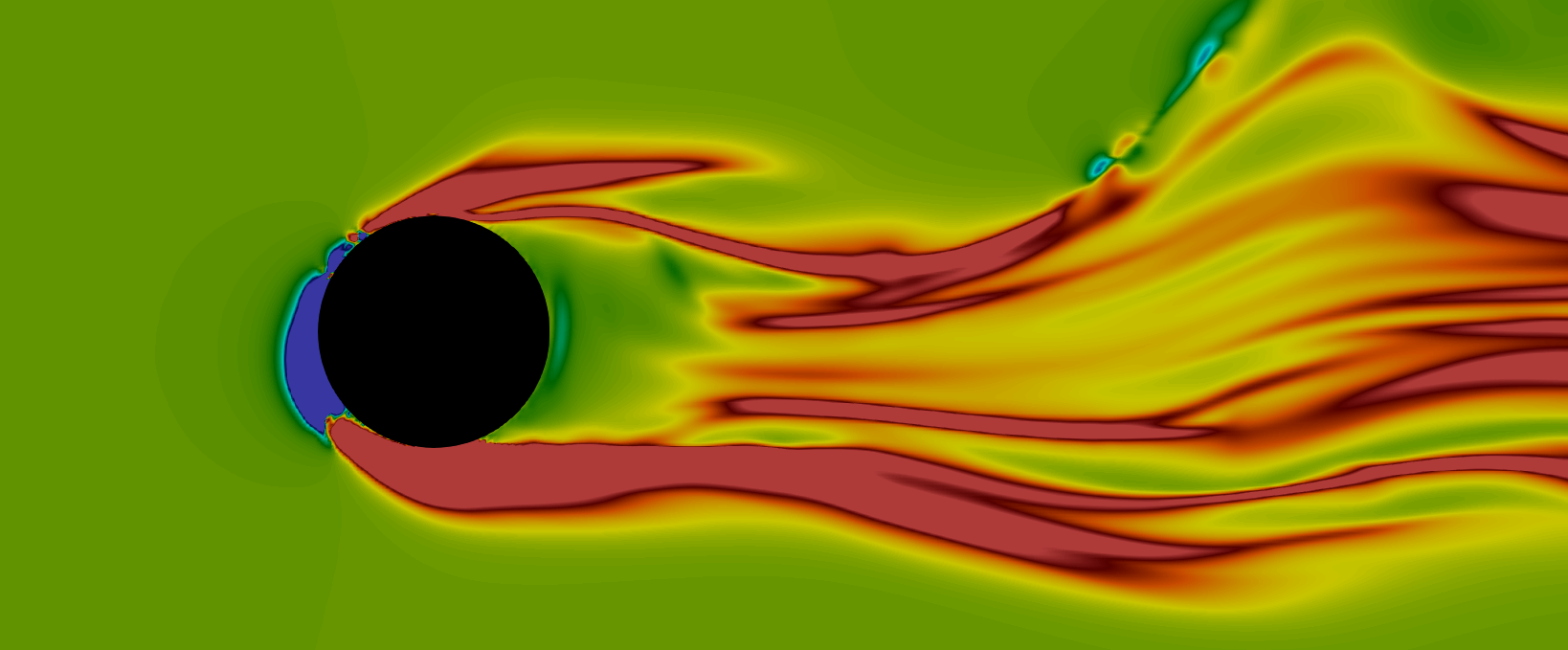}
\caption{Close-up of \Figref{fig:subim13_5}}
\label{fig:subim13_6}
\end{subfigure}
\caption{The first normal stress difference contour $N_{1}=\tau_{xx}^{p}-\tau_{yy}^{p}$ for variation of elastoviscoplastic fluid with $\beta_{s}=0.9$, $Wi=1.0$, $n=[0.2,1.0,2.0]$, and $Bn=2.0$ at $Re=100$. The color scale is the same for all contours.}
\label{fig:image13}
\end{figure}

\Figref{fig:image13} shows the first normal stress difference $N_{1}$ distribution in EVP fluid. When the  yield stress is strong, \ie $Bn=2$, $N_{1}$ becomes larger and thicker streaks of $N_1$ contours are observed compared to the viscoelastic case, as seen in \Figref{fig:image7}. Shear-thinning ($n=0.2$) reduces the magnitude of $N_{1}$ and shear-thickening increases it.
As shown $N_{1} \le 0$ in the vicinity close to the upstream stagnation point due to the change in the flow direction to align to the cylinder surface and it is positive ($N_{1} \ge 0$) for the rest of the domain. Therefore, $\tau_{xx}^{p} \geq \tau_{yy}^{p}$ in the entire domain except close to the upstream stagnation. The local maximum of $N_{1}$ still exists at the outer layer of the vortex at a large enough distance downstream of the cylinder, as seen in \Figref{fig:image13}.

\begin{figure} 
\begin{subfigure}{0.5\textwidth}
\centering
\includegraphics[width=1.0\linewidth, height=2.75cm]{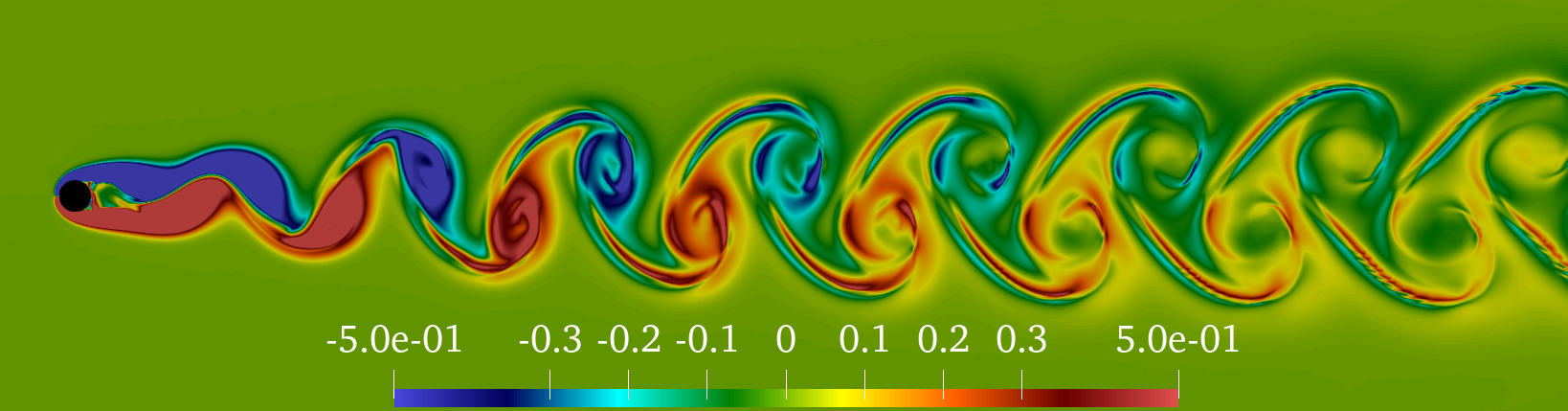}
\caption{\label{Vor_Scale_Re100_HB_Wi1.0_n0.2_Bs0.9_Bn2.0} $Wi=1.0, \beta_{s}=0.9, n=0.2 , Bn=2.0$ }
\label{fig:subim14_1}
\end{subfigure}
\begin{subfigure}{0.5\textwidth}
\centering
\includegraphics[width=1.0\linewidth, height=2.75cm]{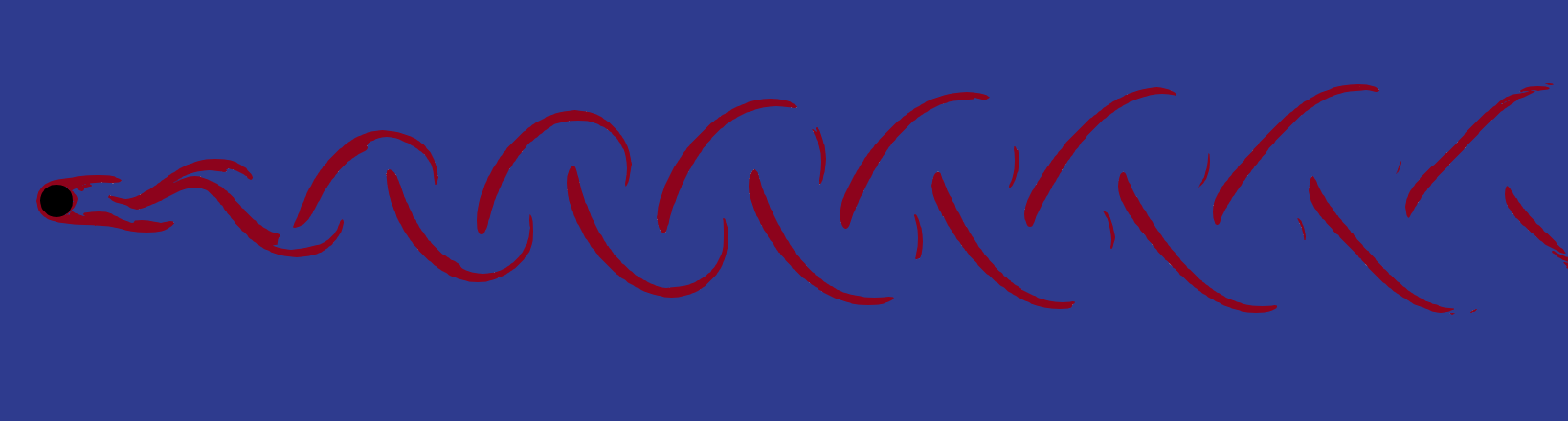}
\caption{\label{T_Scale_Re100_HB_Wi1.0_n0.2_Bs0.9_Bn2.0} $Wi=1.0, \beta_{s}=0.9, n=0.2 , Bn=2.0$ }
\label{fig:subim14_2}
\end{subfigure}
\begin{subfigure}{0.5\textwidth}
\centering
\includegraphics[width=1.0\linewidth, height=2.75cm]{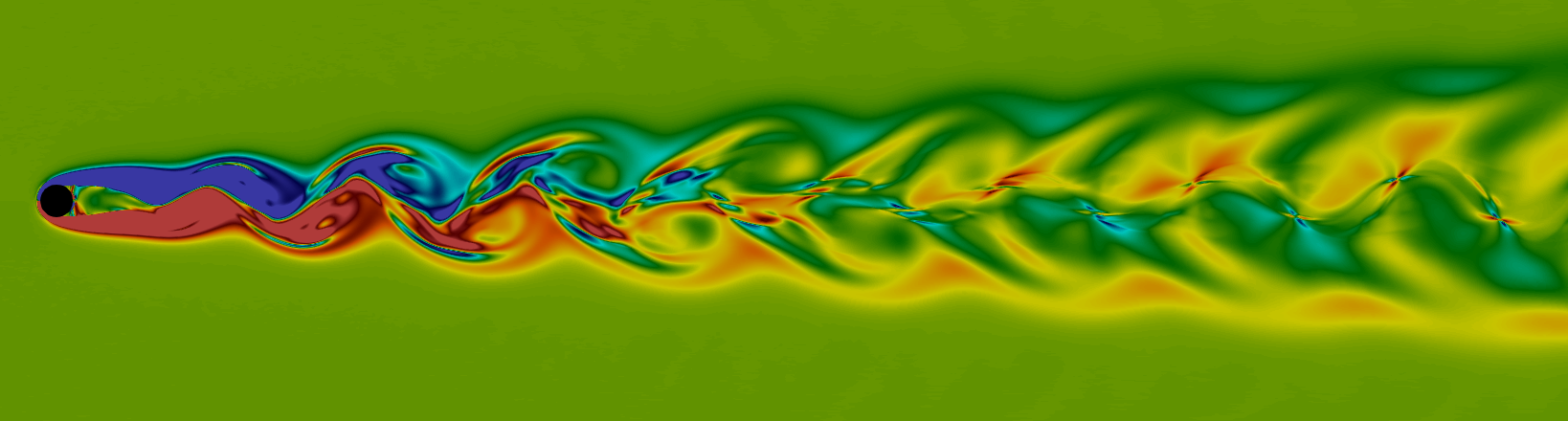}
\caption{\label{Vor_Scale_Re100_HB_Wi1.0_n0.2_Bs0.5_Bn2.0} $Wi=1.0, \beta_{s}=0.5, n=0.2 , Bn=2.0$ }
\label{fig:subim14_3}
\end{subfigure}
\begin{subfigure}{0.5\textwidth}
\centering
\includegraphics[width=1.0\linewidth, height=2.75cm]{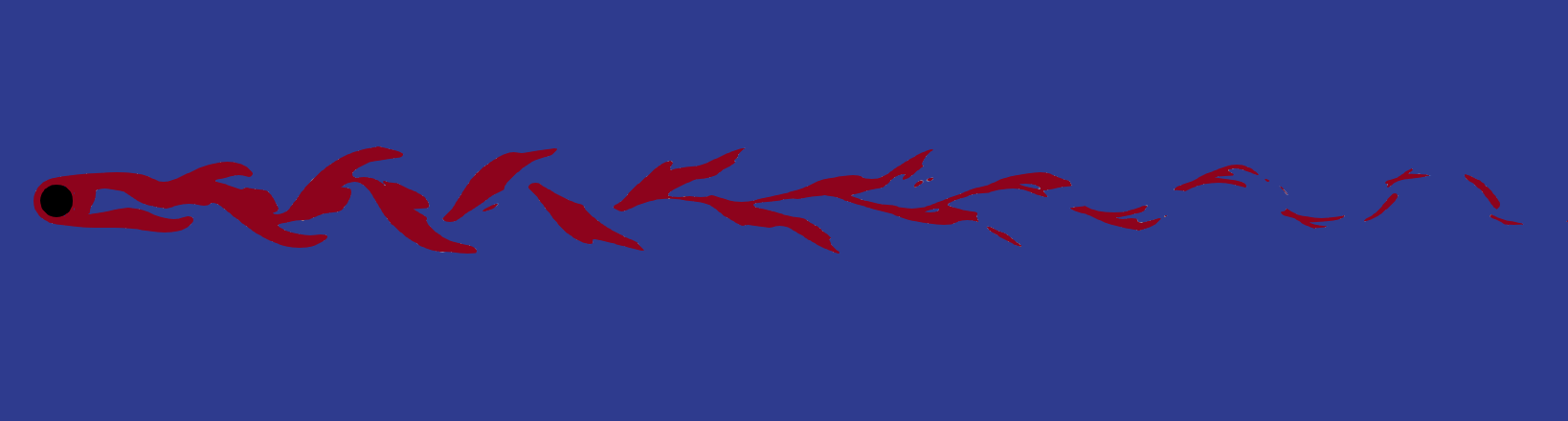}
\caption{\label{T_Scale_Re100_HB_Wi1.0_n0.2_Bs0.5_Bn2.0} $Wi=1.0, \beta_{s}=0.5, n=0.2 , Bn=2.0$ }
\label{fig:subim14_4}
\end{subfigure}
\begin{subfigure}{0.5\textwidth}
\centering
\includegraphics[width=1.0\linewidth, height=2.75cm]{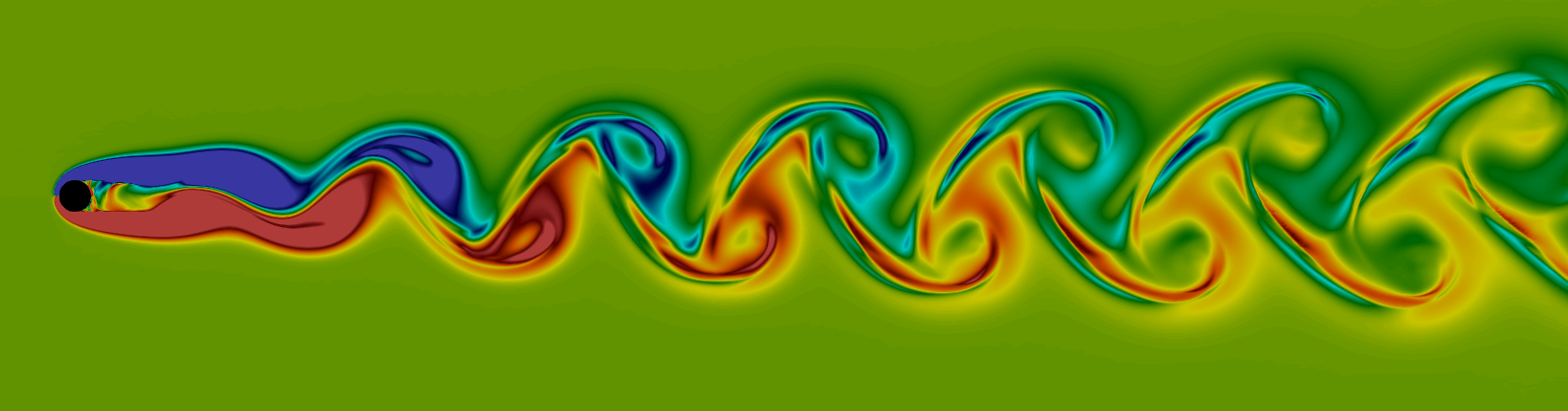}
\caption{\label{Vor_Scale_Re100_HB_Wi1.0_n1.0_Bs0.9_Bn0.0} $Wi=1.0, \beta_{s}=0.9, n=1.0 , Bn=2.0$ }
\label{fig:subim14_5}
\end{subfigure}
\begin{subfigure}{0.5\textwidth}
\centering
\includegraphics[width=1.0\linewidth, height=2.75cm]{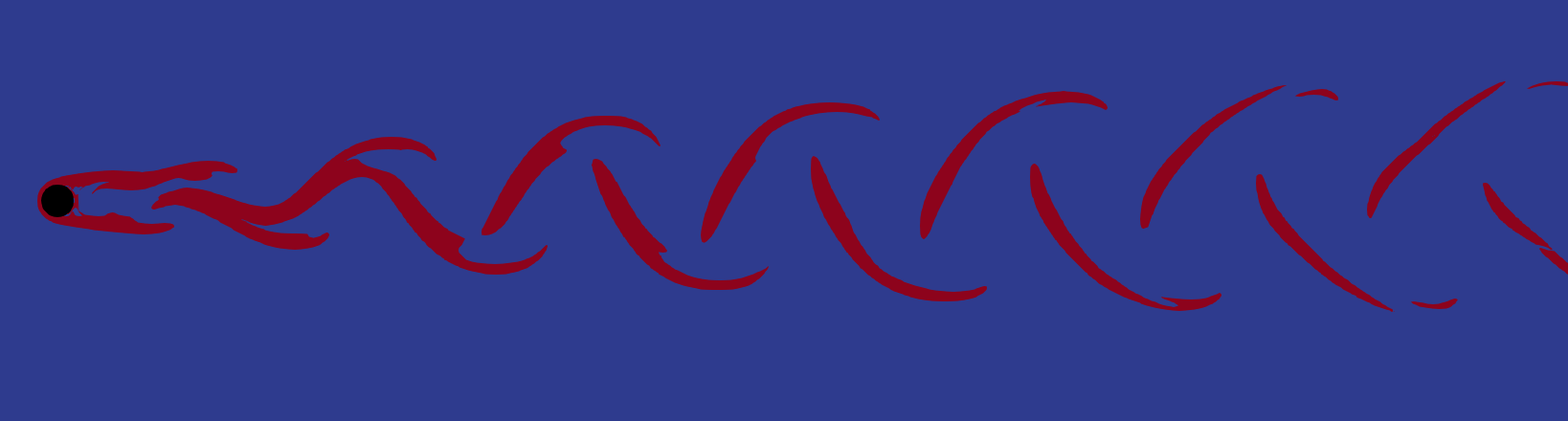}
\caption{\label{T_Scale_Re100_HB_Wi1.0_n1.0_Bs0.9_Bn0.0} $Wi=1.0, \beta_{s}=0.9, n=1.0 , Bn=2.0$ }
\label{fig:subim14_6}
\end{subfigure}
\begin{subfigure}{0.5\textwidth}
\centering
\includegraphics[width=1.0\linewidth, height=2.75cm]{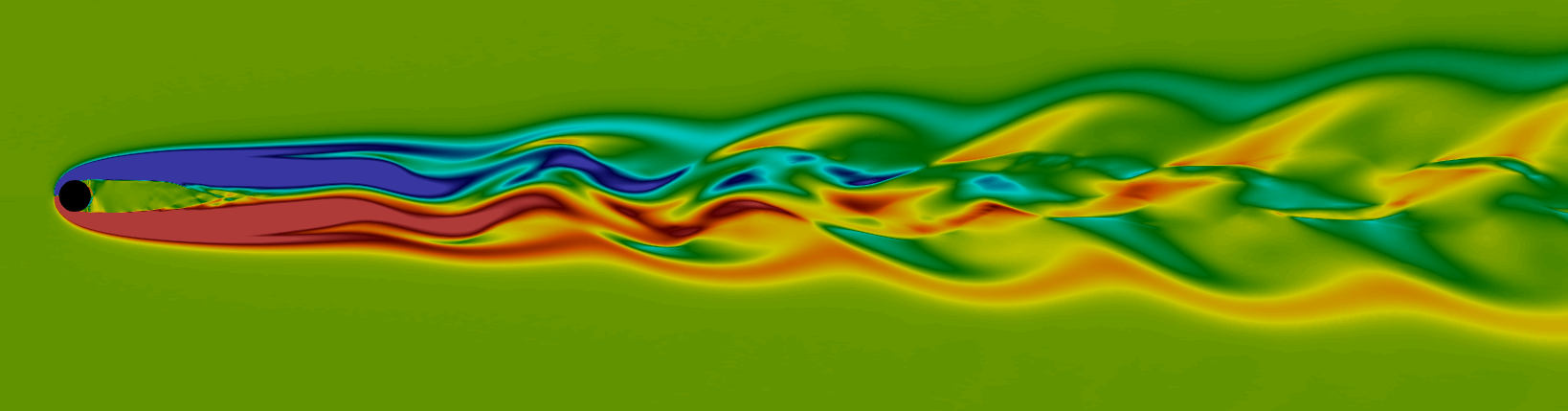}
\caption{\label{Vor_Scale_Re100_HB_Wi1.0_n1.0_Bs0.5_Bn0.0} $Wi=1.0, \beta_{s}=0.5, n=1.0 , Bn=2.0$ }
\label{fig:subim14_7}
\end{subfigure}
\begin{subfigure}{0.5\textwidth}
\centering
\includegraphics[width=1.0\linewidth, height=2.75cm]{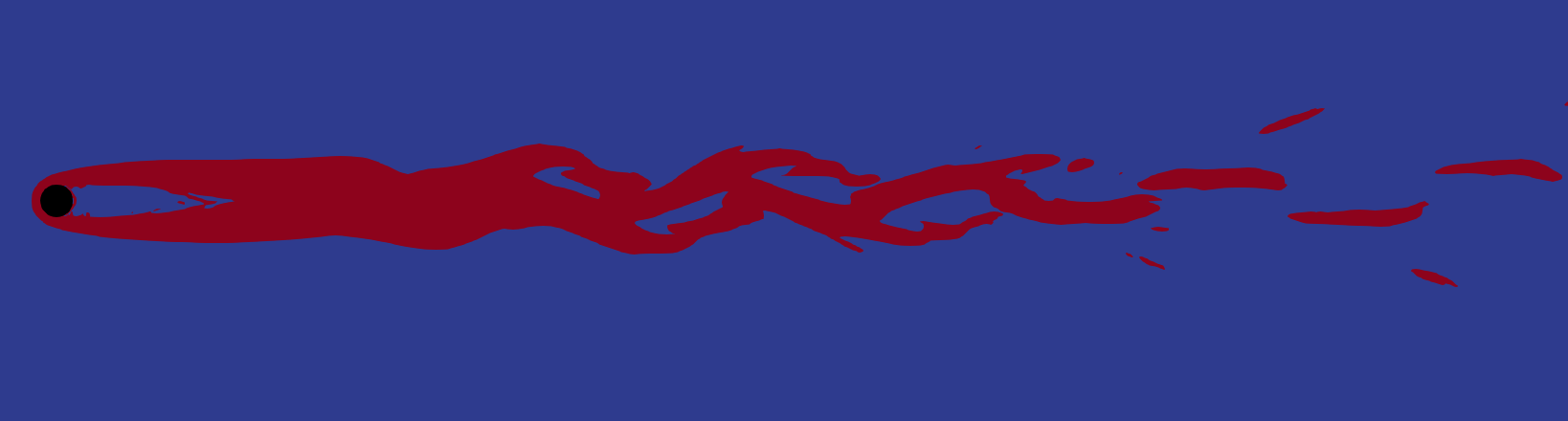}
\caption{\label{T_Scale_Re100_HB_Wi1.0_n1.0_Bs0.5_Bn0.0} $Wi=1.0, \beta_{s}=0.5, n=1.0 , Bn=2.0$ }
\label{fig:subim14_8}
\end{subfigure}
\begin{subfigure}{0.5\textwidth}
\centering
\includegraphics[width=1.0\linewidth, height=2.75cm]{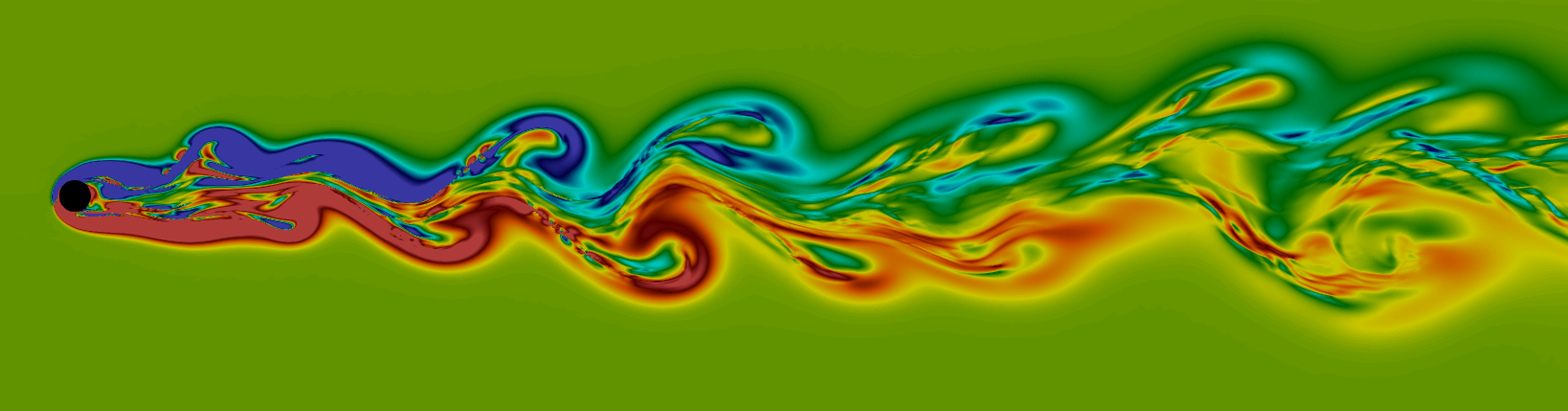}
\caption{\label{Vor_Scale_Re100_HB_Wi1.0_n2.0_Bs0.9_Bn2.0} $Wi=1.0, \beta_{s}=0.9, n=2.0 , Bn=2.0$ }
\label{fig:subim14_9}
\end{subfigure}
\begin{subfigure}{0.5\textwidth}
\centering
\includegraphics[width=1.0\linewidth, height=2.75cm]{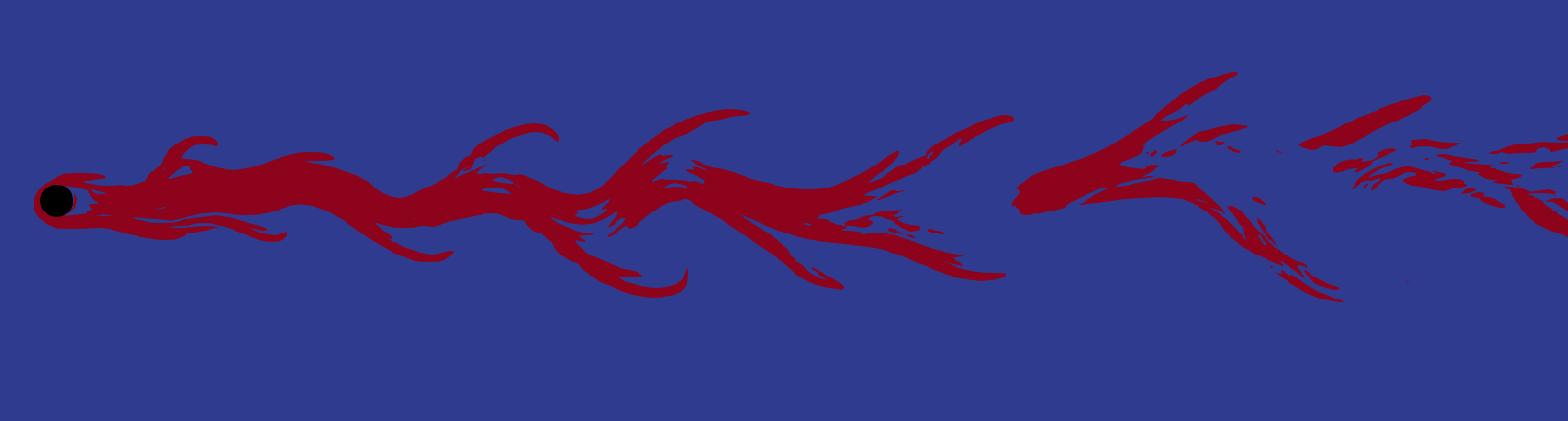}
\caption{\label{T_Scale_Re100_HB_Wi1.0_n2.0_Bs0.9_Bn2.0} $Wi=1.0, \beta_{s}=0.9, n=2.0 , Bn=2.0$ }
\label{fig:subim14_10}
\end{subfigure}
\caption{ (Left) Contours of vorticity and (right) yielded and un-yielded zones (red and blue, respectively) for the different fluids indicated in the panel captions. The color scale is the same in all plots.}
\label{fig:image14}
\end{figure}

The vorticity and yield contours for the EVP wake are shown in figure \ref{fig:image14} at $Bn=2$, $Wi=1$, and at varying $\beta_s$ and $n$. In the presence of yield stress, even at this low Weissenberg number, the vortex structure changes compared to the Newtonian case, especially in the near-wake region. 
As shown in figures \ref{fig:subim14_1}, \ref{fig:subim14_3}, for a dilute EVP solution ($\beta_{s}=0.9$), some Haladie shape vortices form close to the cylinder (as for Newtonian and viscoelastic cases depicted in figure \ref{fig:image2} and \ref{fig:image8}); nevertheless, the vortex structure changes further downstream and rainbow lollipop shape vortices appear due to the yield stress. 

In this case, the magnitude of vorticity at the center of the vortices decays rapidly downstream and approaches zero (shown by the light green color in the vorticity contours) and further downstream the vortices gradually change their direction (shown by the yellow color in the vorticity contours).

For a more concentrated solution ($\beta_{s}=0.5$) with $Bn=2$ in both shear-thinning $n=0.2$ (\Figref{fig:subim14_2}) and shear-independent $n=1$ (\Figref{fig:subim14_4}) cases, cyclonic and anticyclonic vortex structures disappear. Instead, the vortices are extended in the streamwise direction, and a cellular structure appears. For the shear-thickening fluid with $n=2$, the flow pattern becomes chaotic in both dilute and concentrated EVP solutions, so the cellular structure of the vortices also collapses. Generally speaking, irregular velocity fluctuations appear at these elasticity levels when combined with yield stress. These fluctuations are amplified by increasing the power-law index $n$, as the flow becomes shear-thickening. The effect of these fluctuations is not only limited to the area close to the cylinder but also alters the flow structure downstream of the cylinder.

On the right column of \Figref{fig:image14}, contours of the yielded region for the left column cases are depicted. 
The yielded and un-yielded zones in the fluid are distinguished through the following condition: $F = \max\left(0,\frac{|\btau_d^p|-\tau_y}{2k|\btau_d^p|^n}\right)^{\frac{1}{n}}\leq \epsilon$, where we have chosen $\epsilon=10^{-6}$ as the numerical threshold.
Outside the red color region in the left column of the \Figref{fig:image14}, the EVP fluid is un-yielded and behaves as a viscoelastic solid. As expected for the periodic pattern, the yielded region exists at the outer edge of the vortex. However, yielded regions occur also close to the centerline of the domain, where the location of the local maximum polymer stress exists for both concentrated shear thinning and $n=1$ EVP fluid, as seen in \Figref{fig:subim14_4} and \Figref{fig:subim14_6}.

To gain further insight, we do a statistical binary analysis of yielded and unyielded regions. In particular, we  assume $F^{*}=1.0$ when $\epsilon \ge 10^{-6}$ and $F^{*}=0.0$ otherwise, and compute the time average of $F^{*}$ over a sufficiently long time interval and integrate over the computational domain.
The integral measure $F_{int}$ representing the mean yielded region in the flow is made non-dimensional with the area of the computational domain, $L_{x} \times L_{y}$, so that
\begin{equation}
    F_{int} = \frac{\iint F^{*} \,dx\,dy}{L_{x} \times L_{y}}.
    \label{eq:6.1}
\end{equation}

\begin{figure} 
\begin{subfigure}{0.5\textwidth}
\centering
\includegraphics[width=0.9\linewidth, height=6.0cm]{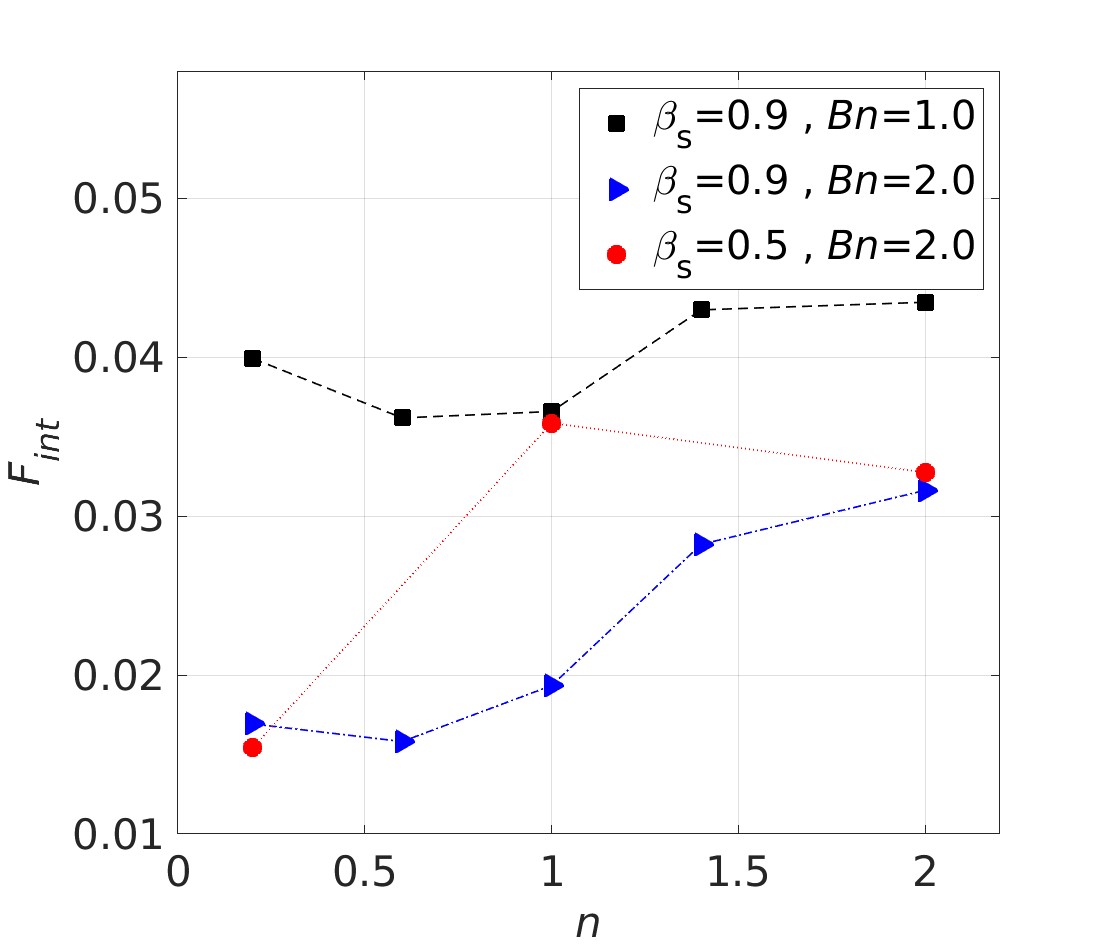}
\caption{\label{Yieled_Region} The yielded Region ($F$):  $Bn=1.0-2.0$}
\label{fig:subim15_1}
\end{subfigure}
\begin{subfigure}{0.5\textwidth}
\centering
\includegraphics[width=0.9\linewidth, height=6.0cm]{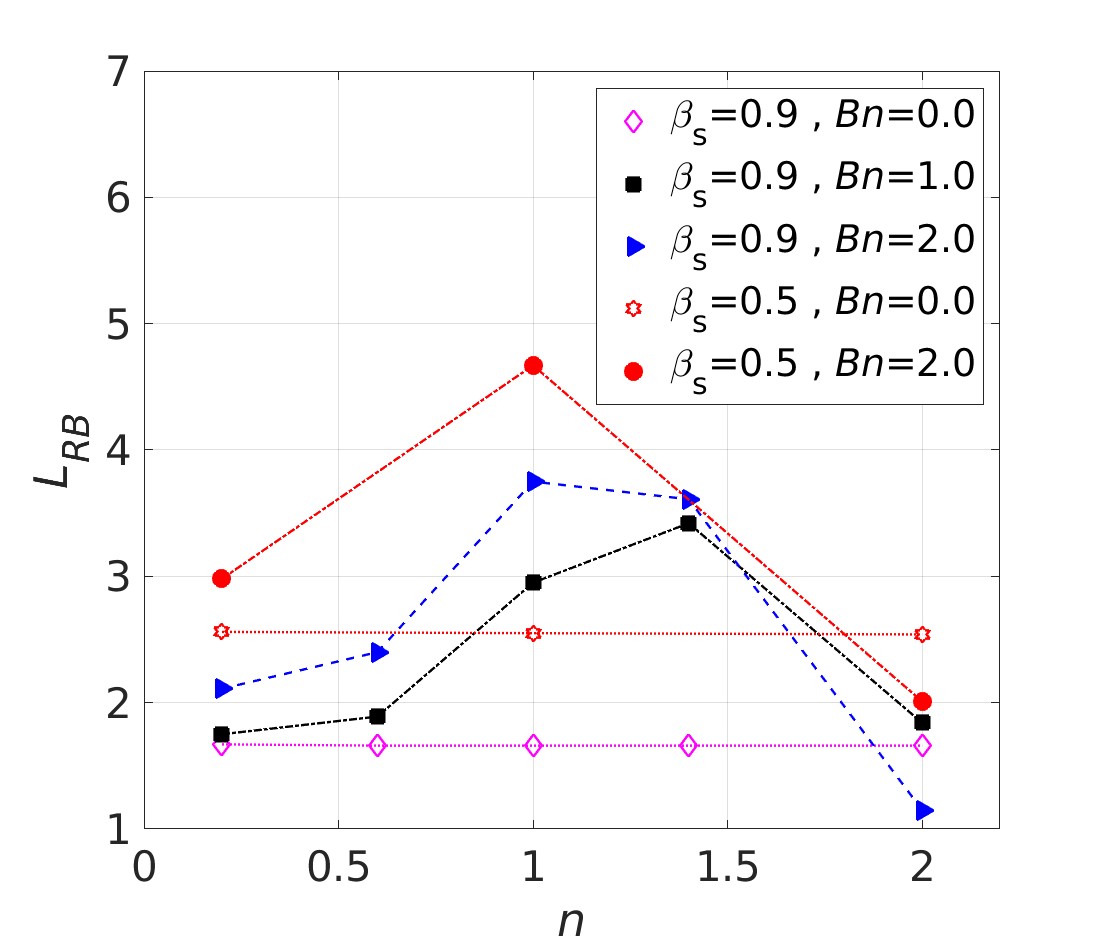}
\caption{\label{EVP_Recirculation} The recirculation length ($L_{RB}$) $Bn=0.0-2.0$}
\label{fig:subim15_2}
\end{subfigure}
\caption{(a) Time average of the percentage of computational domain occupied by yielded fluid (see definition of $F_{int}$ in the text) and (b) recirculation length  ($L_{RB}$) 
versus the power law index $n$, for different values of $Bn$ and $\beta_s$, as indicated in the legend, for the flow of viscoelastic and EVP fluis past a cylinder with $Re=100$ and $Wi=1.0$. In (b)
 viscoelastic flows are indicated by open symbols and EVP flows with filled symbols.}
\label{fig:image15}
\end{figure}

This is displayed in 
\Figref{fig:subim15_1} versus the power law index $n$, for different values of $Bn$ and $\beta_s$, as indicated in the legend. 
For a yield stress such that $Bn=1$ and a dilute solution with $\beta_{s}=0.9$, shear thickening monotonically increases the yielded area, as indicated by $F_{int}$,
while shear-thinning decreases the values of $F_{int}$ when $n=0.6$, but increases again when further decreasing the power index to $n=0.2$. 
For a dilute solution with $Bn=2$, we observe the same trend for the $F_{int}$ variation, but at smaller values, indicating a lower extension of the yielded regions. 
Generally speaking, except for extreme shear-thinning conditions, i.e.\ $n=0.2$, $F_{int}$ increases by increasing $n$. At $\beta_{s}=0.5$, $F_{int}$ considerably decreases in shear thinning fluid and slightly decreased for shear-thickening fluids, unlike the cases with $\beta_s=0.9$. This difference is related to the change from periodic to chaotic time dynamics, which results in a different wake structures at $\beta_s=0.5$, compared to the $\beta_s=0.9$ case.

Finally, \Figref{fig:subim15_2} shows the variations of the recirculation bubble length $L_{RB}$ by the combined effects of elasticity, yield stress, and shear-thinning/shear-thickening.
The data indicate that for $n=1$ and nonzero Bingham number $Bn>0$, 
the length of the recirculation bubble increases considerably when compared with a Newtonian fluid and Oldroyd-B fluid. The presence of yield stress lengthens the recirculation bubble in all cases, except in the chaotic flow at $n=2$. This might be a manifestation that yield stress increases the elastic effects, since $L_{RB}$ also increases with $Wi$, as shown in figure \ref{fig:image5}.
Moreover, shear-thinning ($n=0.6, 0.2$) shortens the length of the recirculation bubble, which could be related to a nonlinear effect on the mean flow by the stronger instabilities at low and high $n$, as evidenced by the higher lift coefficients $C_{L,rms}$ at large and small values of $n$ (see figure \ref{fig:image5}).\Figref{fig:subim15_2} also reveals that shear-thickening has a complicated effect on the length of the recirculation bubble and the location of the recirculation center, as the flow changes from periodic to chaotic. For $n=1.4$, the length of the bubble increases, and its center shifts further downstream. However, for $n=2$, increasing $Bn$ decreases the length of the bubble and causes the center of the recirculation region to shift toward the cylinder. It is worth mentioning that the $L_{RB}$ reduces when the flow pattern becomes chaotic.

It is also observed that in the EVP fluid, the center of the recirculation region shifts towards the cylinder by shear-thinning, while increasing $Bn$ and $n$ or decreasing $\beta_{s}$ cause the opposite effect, not shown for brevity. For a more concentrated solution with $\beta_{s}=0.5$, the recirculation region extends in the streamwise direction for both shear-thinning and shear-thickening flow.

\section{Conclusion}\label{section:Conclusion}
In the present work, the combined effect of yield stress and elasticity has been studied for the flow past a circular cylinder at Reynolds number $Re=100$ by means of direct numerical simulations, in the presence of shear-thinning, shear-independent and shear-thickening viscosity. The cylinder, with no-slip and no-penetration conditions at its surface, is 
constructed by the immersed boundary method, whereas the Saramito Herschel-Bulkley model - in the log conformation tensor formulation - is used to model the rheological behaviour of an elastoviscoplastic fluid with shear-rate dependent viscosity.

In the presence of yield stress, we show that most elastic effects become stronger. The maximum value and downstream extent of trace of the EVP stress tensor increases considerably with $Bn$, and the length of the mean flow recirculation bubble also increases with $Bn$, just like it increases with $Wi$. At higher Bingham numbers, $Bn=1.4-2$, there is no dominant frequency anymore, but flow becomes chaotic and the structures reminiscent of elastic turbulence, despite the low elasticity number. For lower $Bn$, the structure and frequency of vortex shedding is qualitatively similar to the Newtonian case, even though we observe differences in both structure and wavelength of the vortices. 

Apart from yield stress and elasticity, also the shear-rate exponent $n$ has a large effect on the flow. The shear-thinning cases have a very clear dominant frequency with a Newtonian-like vortex shedding, while shear-thickening leads to changes in the wake structure and finally a chaotic flow. Furthermore, the elastic stresses  are weaker in the presence of shear-thinning and stronger in shear-independent and shear-thickening cases. In general, shear-thinning competes with elastic and plastic forces, trying to weaken their effects, and shear-thickening does the opposite. 

Regarding yielded regions, the fluid yields of course around the cylinder, and on the outer side of the shed vortices far downstream, which affects the detailed vortex structure. However, yielded regions occur also close to the centerline of the domain, where the location of the local maximum EVP stress exists for both concentrated shear-thinning and shear-independent EVP fluid.

At a lower solvent-to-polymeric viscosity ratio $\beta_s=0.5$ with $Bn = 2$, we see yet a different wake structure, as the typical cyclonic and anticyclonic vortex structures characteristic of von Kármán vortex shedding disappear. Instead,the vortices are extended in the streamwise direction, and a cellular structure appears.

To summarise, we have found that yield stress enhances the elasticity effect for the flow past a cylinder at moderate Reynolds numbers. The flow pattern (periodic or chaotic) and wake structure changes with the Bingham number, but also depends on several other non-Newtonian parameters. These findings shed new light on the interplay of inertia, yield stress and elasticity on flows behind obstacles, and can serve as a starting point for future experimental studies, and designs of control strategies for non-Newtonian flows.

\section{Acknowledgment}
This work received funding from the European Research Council (Grant no. ERC-2019-StG-852529, MUCUS), and by the Swedish Research Council grant No. VR2021-04820. The authors acknowledge computer time provided by SNIC (Swedish National Infrastructure for Computing)

\bibliography{mybibfile}

\end{document}